\documentclass[10pt,journal]{IEEEtranTCOM}

\usepackage[english]{babel}
\usepackage[usenames]{color}
\usepackage[cp1250]{inputenc}
\usepackage{amsfonts}
\usepackage{amsthm}
\usepackage{graphicx}
\usepackage{epsfig}
\usepackage{mathrsfs}
\usepackage{amsmath}
\usepackage{algorithm}
\usepackage{algorithmic}
\usepackage{hyperref}

\theoremstyle{plain}
\newtheorem{df}{Definition}

\begin{document}
\newcommand{\bea}{\begin{eqnarray}}
\newcommand{\eea}{\end{eqnarray}}
\newcommand{\be}{\begin{equation}}
\newcommand{\ee}{\end{equation}}
\newcommand{\beas}{\begin{eqnarray*}}
\newcommand{\eeas}{\end{eqnarray*}}
\newcommand{\bs}{\backslash}
\newcommand{\bc}{\begin{center}}
\newcommand{\ec}{\end{center}}
\def\SC {\mathscr{C}}

\title{P?=NP as minimization of degree 4 polynomial,\\
integration or Grassmann number problem, \\ and new graph isomorphism problem approaches}
\author{\IEEEauthorblockN{Jarek Duda}\\
\IEEEauthorblockA{Jagiellonian University,
Golebia 24, 31-007 Krakow, Poland,
Email: \emph{dudajar@gmail.com}}}
\maketitle

\begin{abstract}
While the P vs NP problem is mainly approached form the point of view of discrete mathematics, this paper proposes reformulations into the field of abstract algebra, geometry, fourier analysis and of continuous global optimization - which advanced tools might bring new perspectives and approaches for this question. The first one is equivalence of satisfaction of 3-SAT problem with the question of reaching zero of a nonnegative degree 4 multivariate polynomial (sum of squares), what could be tested from the perspective of algebra by using discriminant. It could be also approached as a continuous global optimization problem inside $[0,1]^n$, for example in physical realizations like adiabatic quantum computers. However, the number of local minima usually grows exponentially. Reducing to degree 2 polynomial plus constraints of being in $\{0,1\}^n$, we get geometric formulations as the question if plane or sphere intersects with $\{0,1\}^n$. There will be also presented some non-standard perspectives for the Subset-Sum, like through convergence of a series, or zeroing of $\int_0^{2\pi} \prod_i \cos(\varphi k_i) d\varphi $ fourier-type integral for some natural $k_i$. The last discussed approach is using anti-commuting Grassmann numbers $\theta_i$, making $(A \cdot \textrm{diag}(\theta_i))^n$ nonzero only if $A$ has a Hamilton cycle. Hence, the P$\ne$NP assumption implies exponential growth of matrix representation of Grassmann numbers. There will be also discussed a looking promising algebraic/geometric approach to the graph isomorphism problem - tested to successfully distinguish strongly regular graphs with up to 29 vertices.
\end{abstract}
\section{Introduction}
The P versus NP question is a major unsolved problem of computer science. It asks about existence of a polynomial time algorithm for so called NP-complete problems, for which being a solution can be tested in a polynomial time, however, there is not known efficient way to locate the solution in exponentially large set of possibilities. This class contains many problems which can be reformulated (reduced) one into another through a polynomial transformation. Hence, existence of a polynomial time algorithm for one of them would imply polynomial time algorithm for all of them. Additionally, such hypothetical efficient method would endanger most of currently used cryptography.

Some well known representants of this family are: 3-SAT, Hamilton cycle problem, knapsack problem, travelling salesman problem, subgraph isomorphism problem, clique problem, vertex cover problem, independent set problem, subset sum problem, dominating set problem and graph coloring problem. All of them stay in widely understood field of discrete mathematics, like combinatorics, graph theory, logic.

The unsuccessfulness of a half century search for the answer might suggest to try to look out of this relatively homogeneous field - try to apply advances of more distant fields of mathematics, like abstract algebra fluent in working with the ring of polynomials, use properties of multidimensional geometry, or other continuous mathematics including numerical methods perfecting approaches for common problem of continuous global optimization.\\

While there are approaches for reformulation of 3-SAT into continuous constrained optimization~\cite{cont} of a complex formula which has to additionally fulfill some constraints, this article shows that 3-SAT can be reformulated as the question of just reaching zero of a multivariate real nonnegative degree 4 polynomial, and this degree cannot be further reduced.

This reformulation (reduction) allows to place this problem in both global continuous optimization (unconstrained) and abstract algebra. Hence, it allows to translate the complexity of NP-complete problems to possibly exponential growth of the number of local minima of such multivariate polynomial, what suggests similar issue for recently popular adiabatic quantum computers also representing combinatorial tasks as (energy) optimization problems - might require exponential reduction of temperature to even distinguish between exponentially growing number of local minima.

This degree of polynomial can be reduced further to 2 if we additionally enforce variables to obtain boolean values $\{0,1\}^N$. Zero of degree 2 polynomial is simple linear plane, allowing to polynomially reduce 3-SAT to a geometric problem of plane or sphere intersecting with $\{0,1\}^N$.

Alternatively, from the abstract algebra point of view, polynomial formulation allows to shift the difficulty for example into the problem of calculating multivariate analogue of discriminant of polynomial.

There will be also presented solution of Hamilton cycle problem using anti-commuting Grassmann numbers, suggesting requirement of exponential growth of representation for this algebra. Alternatively, derivatives can be also used for a related approach.

\section{3-SAT as global minimization of polynomial}
3-SAT is the problem of determining if we can assign 0/1 values to boolean variables, such that all alternatives from a chosen set of clauses (triples) are satisfied: $\forall_{i=1,\ldots,m}\ C_i$ . The alternatives may contain negation, for example $C_1\wedge C_2=(x\vee \neg y \vee z)\wedge(\neg x \vee y \vee u)$. Denote $n$ as the number of variables, $m$ as the number of clauses.

We will now translate this conjugation of alternatives into a nonnegative polynomial which zeros correspond to satisfying variable assignments. The boolean variables will be transformed into real continuous variables, which are enforced to finally choose 0 or 1 by the condition of zeroing the polynomial. While the final values have to be discrete, their search might involve intermediately using real values - especially from the $[0,1]^n$ hypercube.

\subsection{Degree 14 polynomial}
The original author's approach~\cite{min} from 2010 has used degree 14 polynomial, reduced to 8 by introducing one additional variable per clause (triple).

Specifically, the $C= x \vee y$ alternative is satisfied in 3 cases: 01, 10 and 11. It is equivalent to zeroing of degree 6 nonnegative polynomial:

$$ p^\vee_2(C):=\left((x-1)^2+y^2\right)\cdot$$
\be\cdot\left((x-1)^2+(y-1)^2\right)\cdot\left(x^2+(y-1)^2\right)\ee
Alternative of three variables is satisfied in 7 cases: 001, 010, 100, 011, 101, 110, 111. Analogously we get degree $7\cdot 2=14$ nonnegative polynomial $p^\vee_3 (C)$, which zeroes if and only if the alternative $C$ is satisfied.

We can now construct the final polynomial as sum of $p^\vee_3$ for all $m$ clauses:
\be p(x_1,\ldots,x_n)=\sum_{i=1,\ldots,m} p^\vee_3(C_i)\ee
where for the negated variables we use $1-x$ instead of $x$. This nonnegative polynomial is zero if and only if all $p^\vee_3$ are zero, what is equivalent with all alternatives being satisfied.

We got degree 14 polynomial of $n$ variables - there is a natural question if this degree can be reduced at cost of at most polynomial growth of the number of variables.

\subsection{Reduction to degree 8}
The original reduction has used additional variables (one per clause) to reduce the number of possibilities satisfying alternative from 7 to 4, hence reducing the degree of polynomial from 14 to 8.

For this purpose, for each $C=x\vee y\vee z$ clause introduce variable $v$ and replace $C$ with conjunction of the following two alternatives:

$$\left((v \wedge (x\vee y)) \vee (\neg v \wedge \neg (x \vee y))\right) \wedge (v \vee z) $$

\noindent The first one looks at four possibilities for $x$ and $y$ variables, enforcing the use of $v=x\vee y$. Thanks of it, the second alternative becomes equivalent to $x\vee y\vee z$.

The left hand side alternative has 4 possibilities for being satisfied, hence can be transformed into minimization of degree 8 nonnegative polynomial.

Summing such $2m$ polynomials for all $m$ c, we get $n+m$ variable nonnegative polynomial of degree 8, which reaches zero if and only if the 3-SAT can be satisfied.

\subsection{Approach for degree 6}
The main contribution of this paper is alternative approach which directly obtains degree 6 polynomial and can be further reduced to degree 4 by adding $m$ variables.

Specifically, observe that $C=x\vee y\vee z$ is satisfied when the sum of representing 0/1 numbers is in $\{1,2,3\}$, leading to degree 6 polynomial:

\be P^\vee_3(C):=(x+y+z-1)^2(x+y+z-2)^2(x+y+z-3)^2\ee

Reaching zero of this polynomial does not enforce variables being in $\{0,1\}$ yet, but it can be done by additional degree 4 polynomials:

\be p(x_1,\ldots,x_n)=\sum_{i=1,\ldots,m} P^\vee_3(C_i) + \sum_{j=1,\ldots,n} x_j^2 (1-x_j)^2 \ee

This final polynomial of $n$ variables is nonnegative, degree 6, and reaches zero if and only if the 3-SAT is satisfied.

\subsection{The final reduction: degree 4}
To reduce to degree 4 polynomial, observe that the $x+y+z=3$ possibility can be avoided by adding a new variable $v$ - instead of $P^\vee_3$ using polynomial:

$$(x+y+v-1)^2(x+y+v-2)^2 + (z-v)^2(z-v-1)^2$$

For $x=y=0$, the zeroing of the left hand side part enforces $v=1$, for which the right hand side part enforces $z=1$. In the remaining cases, the right hand side part allows for $z$ equal 0 or 1.

Summing the corresponding polynomials for all $m$ clauses with polynomials $x_i^2(1-x_i)^2$ for all original $n$ variables and $m$ additional ones, we get a nonnegative degree 4 polynomial of $n+m$ variables, which reaches zero if and only if the 3-SAT is satisfied.

Observe that if P$\neq$NP, this degree 4 generally cannot be further reduced. Nonnegativity requires the degree to be even, so such hypothetical reduction would need to be to degree 2, which can be minimized in polynomial time. However, in the next section we will see that we can further reduce to degree 2, but with additional constraint of all variables being finally boolean, realized here with the $x_i^2(1-x_i)^2$ degree 4 terms.

\subsection{Algebraic approach: discriminant}
We have transformed the 3-SAT problem into the question of reaching zero of e.g. degree 4 nonzegative polynomial with integer coefficients. Such zero (root) would have to be multiple root and abstract algebra has a tool allowing to test if a polynomial has multiple root: it is equivalent to zeroing of discriminant of this polynomial. For example $ax^2 + bx +c$ quadratic polynomial has double root if and only if its discriminant: $b^2-4ac$ is zero.

However, the situation is much more complex for multivariate polynomials~\cite{discr}. For single variable polynomial $P$, discriminant is resultant of $P$ and its derivative $P'$. Resultant of two polynomials is determinant of Sylvester matrix built of coefficients of the two polynomials, of size being sum of their degrees. Direct application of this method to multivariate polynomial would lead to exponential growth of degree. The question is existence of more efficient methods.

Assuming P$\ne$NP, we can formally conclude that the cost of testing multivariate analogue of zeroing discriminant has to grow at least exponentially with the number of variables, even for degree 4 polynomial. Otherwise, we could solve 3-SAT in polynomial time.

\subsection{Global optimization approach}
We have reformulated a 3-SAT problem into testing if a global minimum (can be more than one) of a nonnegative degree 4 polynomial is zero. A natural approach is using some numerical continuous optimization methods, like gradient descent, or simulated annealing. The knowledge that satisfying final values need to be in $\{0,1\}$ allows to terminate the iteration if approaching a vertex of the hypercube (then just test boolean values as rounded all coordinates), or stabilizing far from it (finding a local minimum). Adding some repulsion between multiple considered solutions would allow to find or approximate polynomial number of local minima in polynomial time. However, the number of local minima of polynomial can generally grow exponentially with the number of variables, for example for the $\sum_i x_i^2 (x_i-1)^2$ polynomial. The question is if in practical problems there will appear exponential number of uninteresting (nonzero) local minima.

Hence, from  the P$\ne$NP assumption we can conclude that the number of nonzero local minima of polynomial obtained from 3-SAT problem has generally an exponential growth. There are known ways to reduce the number of local minima by smoothing a function $f$, for example by adding second derivative like Laplacian: considering $f+\lambda\Delta f$ function for some $\lambda>0$, which should be finally reduced to zero in further iterations, for example analogously to adiabatic evolution of adiabatic quantum computers.

Translation of a difficult combinatorial problem into a global (energy) optimization problem is also the base of adiabatic quantum computers, which were shown to be equivalent to standard quantum computers~\cite{adiab}. However, as for this moment, the author is not aware of polynomial quantum algorithm for NP-complete problems. Additionally, this formulation probably also suffers from the exponential growth of local (energy) minima, which might make them thermally indistinguishable - might require exponential decrease of temperature while growing problem size.

\subsection{Other methods for transforming into global optimization problem}
We have discussed transformation of 3-SAT into global optimization of polynomial. The final degree 4 method required adding $x^2 (1-x)^2$ polynomials to enforce final values being in $\{0,1\}$. Alternative approach is using some monotonous function $f:(-\infty,\infty)\to (0,1)$, for example $f(x)=1/(1+\exp(-x))$ or $f(x)=\arctan(x)/\pi +1/2$, and expect $x\to \pm\infty$ during optimization by using $f(x)$ instead of the original variables in the optimization problem.

While polynomials allow to enforce one of a few possibilities - using polynomial of twice higher degree, in the everyday problem of correcting Low Density Parity Check~\cite{LDPC} error correction codes, the constraints are enforcing parity of all chosen subsets of variables, what can be realized by adding periodic functions like $\sin^2(\pi x )$.

Analogously, periodic functions can be for example used to formulate the problem of integer factorization of $n$ as maximization of
$$\cos(2\pi x) + \cos(2\pi n/x)$$
where $x$ and $n/x$ are the two factors and this sum reaches 2 if and only if both are integer.
\subsection{Example for cryptographic attacks}
While transforming into minimization of degree 4 polynomial will rather not lead to a formal proof of existence of polynomial time algorithm, the more practical question is if it could lead to essentially faster search than brute force among $\{0,1\}^n$ boolean values. Thanks to continuity and simplicity of the resulting polynomial, one could try methods for example based on gradients: which take the search inside the continuous $[0,1]^n$ hypercube and exploit gradients - suggesting local direction to continue the search basing on the entire problem.

Naively, these local suggestions inside $[0,1]^n$ hypercube might lead to essentially faster search than brute force, endangering current cryptography. For example hash functions are designed to be easily propagated in the intended direction of calculation, but seem extremely difficult to propagate in backward direction. In theory, such algorithm of calculation can be written as a Turing machine, which can be transformed into a 3-SAT problem of at most polynomially larger size, which then can be transformed into minimization of degree 4 polynomial as discussed here. In practice, these hash functions are usually constructed as a fixed sequence of basic boolean and arithmetic operations, which can be nearly directly transformed into 3-SAT clauses and polynomials for minimization. For example $z = x\wedge y$ into minimization of $(z-xy)^2$, $z=x\vee y=\neg(\neg x \wedge \neg y) $ into $(z-1+(1-x)(1-y))^2$, $z=x\oplus y$ into $((x-y)^2-z)^2$. Summation of two integers as bit sequences: $z=x+y$ can be transformed into sum over all bit positions $i$ of $(x_i+y_i+c_i - 2c_{i+1} -z_i)^2$ polynomial, where $c_i$ are carry bits.

Thanks to decomposing a hash function into a set of simple basic blocks like 3-SAT clauses, these blocks no longer emphasize any direction of propagation, gradient of the resulting polynomial might be a tool to propagate this information also in the opposite than intended direction. For example fixing the final hash value (bit sequence) and propagating it to the initial value, might allow to reverse this hash function. Fixing some number of leading zeros and part of the input might allow for less expensive Proof-of-Work for example for Bitcoin mining.

Other basic problem which might be attacked through efficient 3-SAT saving is RSA: integer factorization as propagating information from a fixed product of two integers into these two values. Analogously for the discrete logarithm problem being the base of elliptic curve cryptography. For symmetric cryptography, all but the proper key lead to a completely random decoded sequence - we could transform the search for the only key leading to a correlated decoded sequence as a 3-SAT problem.

Hence the essential question is if the local gradient would be a helpful hint for the search? If true, it could lead to cryptographic attacks which are essentially faster than brute force, e.g. by performing gradient descent from multiple random initial points of the $[0,1]$ hypercube, or some more sophisticated numerical optimization, or through some physical realization like adiabatic quantum computer. For protection against such hypothetical attacks, we could elongate the reason-result chain leading to the output to make propagation in opposite direction more difficult.\\

There were performed tests of such approach for the factorization problem: write in binary operations the process of multiplication of two integer numbers in their binary representations, which were finally translated into degree 4 polynomial. Then binary representation of product was fixed as product of some two prime numbers.

There were performed trials to minimize this polynomial to find the two prime factors. Performing gradient descent from many random initial points did not bring essential improvement comparing to brute-force, nor did smoothing with Laplacian. There were also tested methods restricting the polynomial to a line, where we can analytically find all the minima for such low degree polynomial, however, random directions usually contained only a single minimum.

As expected, the found polynomial has huge number of local minima with value close to zero. They are close to $\{0,1\}^n$ hypercube vertices due to $x_i^2(1-x_i)^2$ terms in the sum. Most of these vertices do not solve the problem, hence have a nonzero value of some other terms in the sum. The local minima were perturbations of these vertices such that only one or a few logical operations were violated - such optimization intuitively breaks the logical reason-result chain in some weak links. The number of subsets of terms not to be satisfied grows exponentially, many of them correspond to a local minimum in the extremely complex landscape of this polynomial.

Analogous problem of exponential growth of local minima is expected for physical realizations like adiabatic quantum computers - the experiments suggest the gap between the two lowest states also drops exponentially. Maintaining the lowest state would need not only exponential reductions of speed to have adiabatic process, but also exponential reductions of temperature to make these states them distinguishable.

\section{Geometric formulations}
The degree 4 while transforming into polynomial is required due to $x^2(1-x)^2$ terms enforcing final boolean values. It turns out that the remaining constraints can be represented by degree 2 polynomial. Its zeros are in a plane which can be found in polynomial time.

\subsection{Degree 2 polynomial in vertices of hypercube}
Assume we somehow enforce the final values to be boolean: vertices of hypercube, for example by adding sum of $x_i^2(1-x_i^2)$ over all variables to our polynomial.

In this case, observe that $x\vee y \vee z$ clause from 3-SAT problem can be transformed into minimization of

$$(x+y+z-3u-2v-w)^2 + (u+v+w-1)^2$$

\noindent where $u,v,w$ are additional new variables (3 per clause), which are also somehow enforced to be finally in $\{0,1\}$. The right hand side square acts as alternative: its minimization means exactly one of them is 1, each possibility corresponds to a different $x+y+z\in\{1,2,3\}$ fulfilling the clause. This way we have increased the number of variables from $n$ to $N=n+3m$, where $m$ is the number of clauses.

Here are examples of transforming other basic logical operations ($u,v$ are additional new variables enforced to be boolean):
\begin{itemize}
  \item $x\wedge y$ as $(x+y-2)^2$,
  \item $z=x\wedge y$ as $(x+y-2z-u)^2$,
  \item $z=x\vee y$ as $(x+y-2u-v)^2+(u+v-z)^2$,
  \item $z=x\ \textrm{xor}\ y$ as $(x-y+u-v)^2+(u+v-z)^2$.
\end{itemize}
\subsection{Plane intersecting hypercube vertices problem}
The obtained degree 2 polynomial can be written as \be \frac{1}{2} \textbf{x}^T A' \textbf{x} -\textbf{x}^T \textbf{b}' +a_0 \label{parab}\ee

\noindent Where $\textbf{x}$ is vector of all $N$ variables (e.g. $N=n+3m$ for transformation of 3-SAT above), $A'$ is $N\times N$ matrix with integer coefficients for the above transformation, of absolute values bounded by $O(N)$.

As by construction this polynomial is nonnegative, the $A'$ matrix is positive semi-defined and $a_0\geq 0$. Differentiating (\ref{parab}), it reaches 0 if $a_0=0$ and $A'\textbf{x}=\textbf{b'}$, which is equation of $N-d$ dimensional plane, where $d$ is the order of matrix $A'$ (maximal number of linearly independent rows).

Using Gram-Schmidt orthogonalization ($O(N^3)$ time complexity), we can choose a size $d$ maximal linearly-independent subset of rows of $A'$ - let us construct rectangular $d\times N$ integer matrix $A$ from these rows, and choose $\textbf{b}\in \mathbb{Z}^d$ as the corresponding coordinates of $\textbf{b}'$. Hence $A$ is maximally reduced integer matrix defining the same plane: $\{\textbf{x}:A'\textbf{x}=\textbf{b}'\}=\{\textbf{x}:A\textbf{x}=\textbf{b}\}$.

Finally we have polynomially reduced 3-SAT problem into the following problem:

\begin{df}Plane crossing hypercube vertices problem - decide if $\{0,1\}^N$ intersects a given plane $\{\textbf{x}:A\textbf{x}=\textbf{b}\}$ defined by integer $d\times N$ matrix $A$ and $\textbf{b}\in \mathbb{Z}^d$ vector:
\be \{\textbf{x}:A\textbf{x}=\textbf{b}\}\cap \{0,1\}^N =\emptyset.\ee
\end{df}

As we are interested in $\textbf{x}\in\{0,1\}^N$, the $A\textbf{x}=\textbf{b}$ equation becomes a question of existence of subset of $N$ columns of $A$ which sum to $\textbf{b}$. The coefficients of $A$ are bounded by $O(N)$, hence we can pack the entire columns into large numbers of $\approx d \lg(N)$ bits, getting standard subset sum NP-complete problem: for a given set of integer numbers, is there a subset summing to zero. Hence, we get a geometric analogue of subset sum problem, which may lead to some new approaches.

\subsection{Sphere intersecting hypercube vertices problem}
All $\{0,1\}^N$ are in euclidean sphere with center $\textbf{c}'=(1/2,\ldots,1/2)$ and radius $r'=\sqrt{N}/2$. Intersecting this sphere with the $\{\textbf{x}:A\textbf{x}=\textbf{b}\}$ plane, we get sphere of center in $\textbf{c}$ being orthogonal projection of $\textbf{c}'$ into the plane, and radius $r^2=r'^2-\|\textbf{c}-\textbf{c}'\|^2$.

Moreover, intersection of both spheres lies in the plane of interest. Hence, points of intersection $S(\textbf{c},r)\cap \{0,1\}^N$ are also in the plane, solving e.g. the original 3-SAT problem. Finally we have reduced it to:

\begin{df}Sphere crossing hypercube vertices problem - decide if $\{0,1\}^N$ intersects a given sphere:
\be S(\textbf{c},r)\cap \{0,1\}^N =\emptyset.\ee
\end{df}

Writing the norm as sum over coordinates, the choice among $\{0,1\}^N$ possibilities again becomes equivalent with the subset sum problem, giving it another geometric interpretation. Additionally, beside euclidean sphere here, it could be alternatively a sphere for a different norm - in the next section there is derivation focusing on $l^1$ sphere.

\section{Subset Sum as hyperplane/sphere intersection or integration problem}
The above considerations have lead us close to the known Subset Sum problem, for  which one of formulations is:
\be \exists\  a\in\{0,1\}^{n'}:\ \sum_i a_i x_i = s' \label{ss1} \ee
\noindent for some integer $x_i \in \mathbb{Z}$ defining the instance of problem.

Subtracting $\sum_i x_i/2$ from both sides of the above sum, multiplying both sides by 2, then introducing:

$$ y_i = |x_i| \quad s = 2\left(s'-\frac{1}{2}\sum_i x_i\right) $$

\noindent we get equivalent formulation for natural $y_i\in\mathbb{N}$:

$$ \exists\ a \in\{-1,1\}^{n'}:\ \sum_i a_i y_i = s $$

Appending $s$ to the set of values ($n=n'+1,\ y_n=|s|$) we get equivalent problem for natural $y_i\in\mathbb{N}$:

\be \exists\ a \in\{-1,1\}^n:\ \sum_i a_i y_i = 0 \label{ss2}\ee

So we have a set of natural numbers, and the problem is to split it into two disjoint subsets having the same sum $\left(\sum_i y_i/2\right)$. Alternatively, we get the original Subset Sum problem (\ref{ss1}), but with all values being natural numbers (positive) and the searched sum being  $s'=\sum_i y_i/2$.

\subsection{Plane or sphere intersection problem}
From geometric perspective, the above problem can be seen as the question if hyperplane defined by a normal vector of integer (or natural) coefficients intersects $\{0,1\}^n$ or $\{-1,1\}^n$.\\

To go to the sphere intersection problem, choose some number $d > \max_i y_i$ (natural or real), add it to each term of (\ref{ss2}) sum, then divide its both sides by $d$, getting equivalent condition:
\be \sum_i \left(1 + a_i \frac{y_i}{d}\right) = n \ee
The $a_i=\pm 1$ chooses between two positive values ($d > \max_i y_i$), which sum to 2. It can be seen as choosing between distances from $\mp 1$ points. Hence the original Subset Sum problem is satisfied iff:

\be S_1\left(\left(\frac{y_i}{d}\right)_i, n\right) \cap \{-1,1\}^n \neq \emptyset \label{ss3}\ee
\noindent for $S_1(c,r):=\{x:\ \sum_i |x_i-c_i| =r\}$ is $l^1$ sphere and any $d > \max_i y_i$ (can be also generalized to other $l^p$ spheres).

Condition (\ref{ss3}) can equivalently be seen as finding characterisation of union of spheres:

\be \left(\frac{y_i}{d}\right)_i\ \in\ \bigcup_{c\in  \{-1,1\}^n} S_1(c,n)\ =\ S_1(0,n) + \{-1,1\}^n \ee

\begin{figure}[t!]
    \centering
        \includegraphics{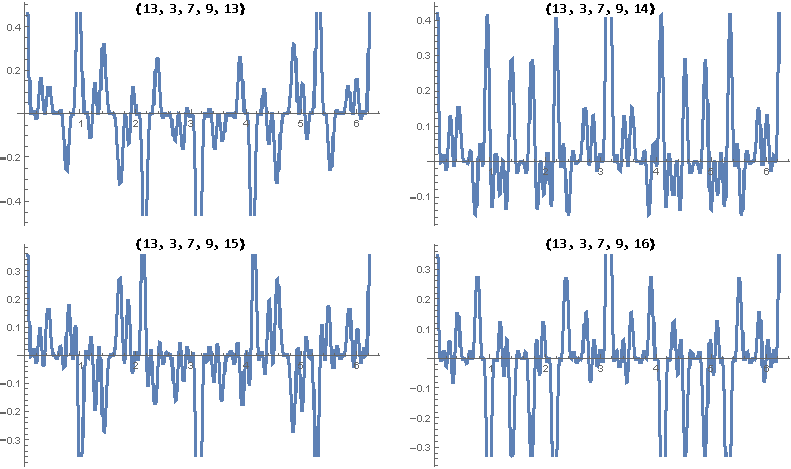}
        \caption{Four examples of $\prod_i \cos(\varphi\, y_i)$ functions on $[0,2\pi]$ range for the written $\{y_i\}_{i=1..5}$, which differ only by $y_5$ here. As in (\ref{intfor}), this plot has nonzero average iff $\{\sum_i a_i y_i:\ a_i\in\{-1,1\}\}$ contains zero. Only the upper-right plot fulfills this condition.}
       \label{integrate}
\end{figure}

\subsection{Integration formulation}
Another interesting formulation of the Subset Sum problem is by calculating Fourier transform of the characteristic function (with multiplicities) of all possible $2^n$ values:
\be X:= \left\{\sum_i a_i y_i:\ a_i\in\{-1,1\}\right\} \ee
$$\sum_{x\in X} e^{\varphi x\textbf{j}}=\prod_i \left(e^{\varphi y_i\textbf{j}} +e^{-\varphi y_i\textbf{j}}\right) = 2^n \prod_i \cos(\varphi y_i) $$

Without the $2^n$ term it is just the probabilistic characteristic function for uniform probability distribution on $X$ (with multiplicities): sum of independent random binary variables $\Pr(X_i=y_i)=\Pr(X_i=-y_i)=1/2$.

The original question if the $X$ contains 0 becomes the question if:
\be 0 \neq \int_0^{2\pi} \prod_i \cos(\varphi\, y_i)\, d\varphi \label{intfor}\ee

It transforms to the original problem when expressing cosines as complex exponents. Also if treating it as integration over complex unit circle to use the residue theorem - testing if 0 is a residue becomes again the original problem:

$$0\neq \oint z^{-1}\, dz\, \prod_i (z^{y_i} + z^{-y_i}) $$

Different approaches for integrating $\prod_i \cos(\varphi\, y_i)$, like averaging over random points (Monte Carlo) seem also inefficient because, while this product of cosines has values in $[-1,1]$, the nonzero average (integral) drops pessimistically like $2^{-n}$.
\begin{figure*}[t!]
    \centering
        \includegraphics{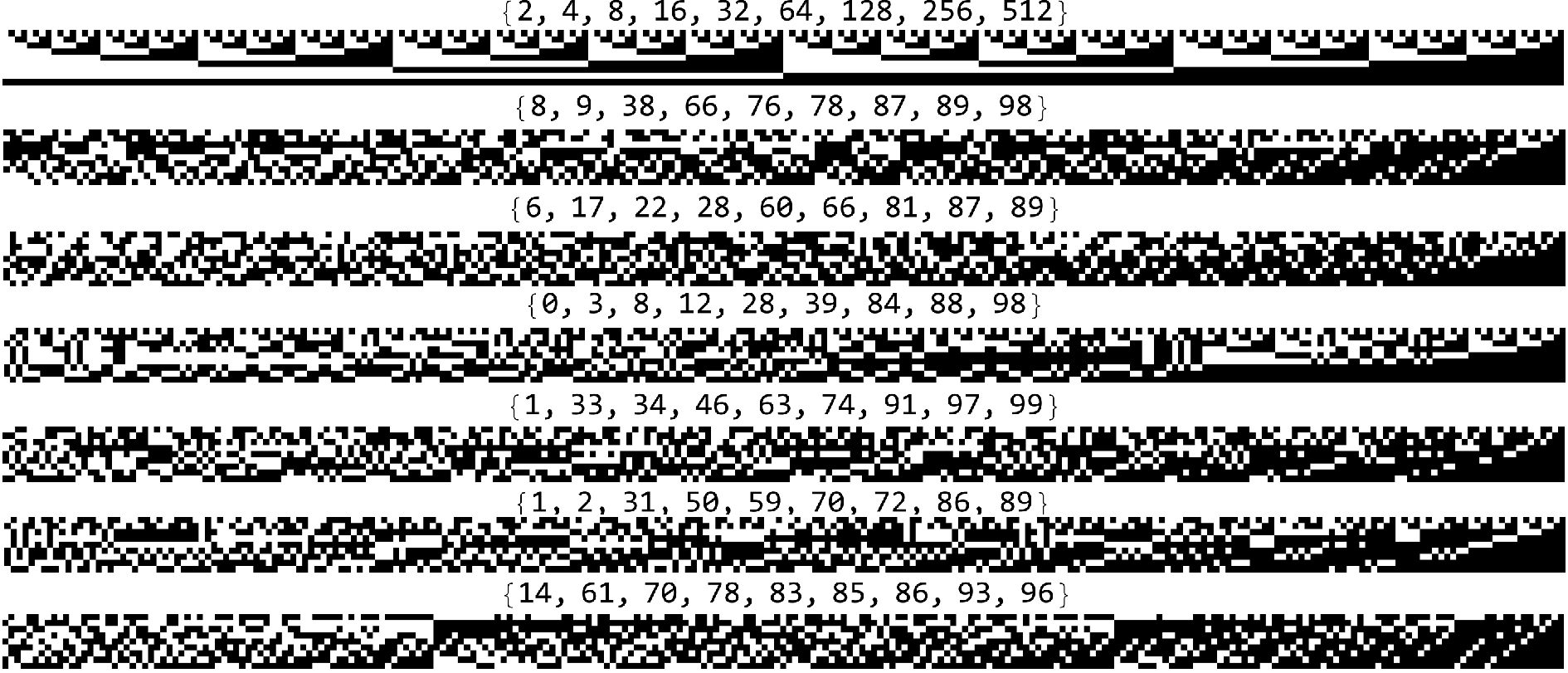}
        \caption{The order of positive values in $X=\{\sum_i a_i y_i:\ a_i\in\{-1,1\}\}$ set for some different $\{y_i\}_{i=1..9}$ (written). Black squares correspond to $a_i=+1$, white to $a_i=-1$. Obviously, the largest values have all $+1$, what can be seen as black part on the right. Generally the pattern is quite complicated. The question is if some regularity can be found to quickly point suspects for summing to zero?}
       \label{ssorder}
\end{figure*}

A related approach is looking at $\prod_i (z^{y_i}+z^{-y_i})$ as Laurent series and asking for $z^0$ coefficient, or at $\prod_i (1+z^{x_i})$ polynomial and asking for coefficient of $z^s$. Doing it by differentiation takes us back to the original problem.
\subsection{Symmetric polynomials approach}
While the original question if the set of $2^n$ possibilities: $X= \{\sum_i a_i y_i:\ a_i\in\{-1,1\}\}$ contains zero seems very difficult (NP), surprisingly, we can inexpensively calculate sums of small natural $p$-th powers of all these $2^n$ values:
\be t_k :=\sum_{x\in X} x^p\qquad\qquad s_k:=\sum_i y_i^p \ee
Obviously $t_p=0$ for all odd $p$. For even $p$, $t_0=2^n$,
$$t_2=\sum_{a\in\{-1,1\}^n} \left(\sum_i a_i y_i \right)^p =2^n s_2$$
\noindent thanks to cancellations due to alternating signs: the only non-vanishing terms are those having even numbers of all appearances while expanding the power. The relation for higher powers can be analogously found (Mathematica source below), but it becomes more complicated, for example:
$$t_4 = 2^n(3s_2^2-2s_4)$$
$$t_6=2^n(16 s_6-30s_2 s_4 +15 s_2^3)$$
$$t_8=2^n(-272s_8 +448 s_2 s_6+140 s_4^2 -420 s_2^2 s_4 +105 s_2^4)$$

We see formulas similar to those relating cumulants with central moments. This is not a coincidence as uniform probability distribution on $X$ (with multiplicities) is sum of binary random variables. Cumulants of $i$-th variable are coefficients of Maclaurin series of $\ln((e^{t y_i}+e^{-t y_i})/2)$, and they are additive for independent random variables, so cumulants for $X$ are sums of the original ones. The question is if we can effectively use these cumulants to find probability of zero? Or expected value of a function divergent in 0 like $1/x^2$ (or e.g. $\ln(x)=\lim_{n\to 0} (x^n-1)/n$ replica trick), for which analytic formula would allow to test if $0\in X$.

The original problem can be alternatively formulated as the question if product of such $2^n$ values is zero. Product of all values is also a symmetric polynomial and could be expressed as a polynomial of the above values (e.g. $2xy=(x+y)^2-(x^2+y^2)$), but it would require  $2^n$ of them.\\

Below is Mathematica source calculating the above coefficients, the big question is if it could be simplified to inexpensively test e.g. if $\sum_{x\in X} x^{-2}$ is finite, e.g. through some Taylor expansion.
\begin{scriptsize}
\begin{verbatim}
maxp = 14; cd = {{{1}}}; divs = {cd};       (* find divisions *)
Do[AppendTo[divs, cd = Flatten[ Table[Append[
   Table[cc = c; cc[[i]]=Append[cc[[i]], k]; cc, {i,Length[c]}],
     Append[c, {k}]], {c, cd}], 1]], {k, 2, maxp/2}]; 
Do[fp = Factorial[p];   (* calculate coefficients *)
 part = Map[Sort,2 IntegerPartitions[p/2]];(* even partitions *)
 kp = Table[
  fp/Apply[Times,Factorial[Join[s,Values[Counts[s]]]]],{s,part}];
 mx = Flatten[Table[cp = part[[i]]; cd = divs[[Length[cp]]]; 
  cn=Counts[Table[Sort[Table[Total[cp[[cc]]],{cc, c}]],{c, cd}]];
  ks = Flatten[Table[Position[part, c], {c, Keys[cn]}]];
  Table[{i, ks[[j]]} -> Values[cn][[j]], {j, Length[ks]}]
    ,{i, Length[part]}], 1];
 {kp . Inverse[SparseArray[mx]], part} // MatrixForm // Print 
   ,{p, 2, maxp, 2}]
\end{verbatim}
\end{scriptsize}
\subsection{Convergence approach}
Instead of summing a natural power of all $2^n$ values, we could for example use a function being infinity in zero and try to test is the sum is infinity. For example
\be T:=\sum_{a\in \{-1,1\}^n}\frac{1}{\left(\sum_i a_i y_i\right)^2}=\sum_{a\in \{-1,1\}^n}\frac{1}{s_2+\sum_{i\neq j} a_i a_j y_i y_j} \ee
\noindent where $s_2=\sum_i y_i^2$. Expanding the series:
$$ s_2 T = \sum_{k=0}^{\infty} \sum_{a\in \{-1,1\}^n} \left(-s_2^{-1}\,  \sum_{i\neq j} a_i a_j y_i y_j \right)^k $$

\noindent Unfortunately cancellation due to alternating signs is much more complex this time, making problematic calculation for high $k$. From the other side, e.g. using a series of $t_k$ above, divergence to test is relatively slow: $1/(1-1)=\sum_k 1$, requiring very large orders to distinguish zero value for one close to zero in such sum.\\

A similar trial can be made for different mentioned formulation: finding subset summing to $s=\sum_i y_i/2$:
$$\sum_{a\in \{0,1\}^n}\frac{1}{s - \sum_i a_i y_i}=\frac{1}{s}\sum_{k=0}^{\infty} \sum_{a\in \{0,1\}^n}\left(\frac{1}{s}\sum_i a_i y_i \right)^k $$
\section{Some further examples}
Let us briefly look at some further examples of converting NP-complete problems into algebraic, optimization or algebraic problems.
\subsection{Optimization formulation of clique problem}
In clique problem we ask if a given graph contains a size $k$ clique. If $A$ is the adjacency matrix, clique problem corresponds to maximizing:

$$\max_{v\in \{0,1\}^n} \left\{v^T A v:\ \sum_i v_i =k\right\}. $$

\noindent Which reaches maximal value $(k-1)^2$ only for clique. Equivalently, we can search for anti-clique: ask if
$$0\in\left\{v^T A v:\ \sum_i v_i =k,\ v\in \{0,1\}^n\right\}\ ?$$
Taking kernel of $A$, we again get the question of linear subspace crossing vertices of hypercube.

\subsection{Vertex cover using cone counterimage}
For vertex cover problem we get question if the following set is nonempty (assume here $\forall_i\,A_{ii}=1$):
$$\left\{Av\geq \mathbf{1}:\ \sum_i v_i =k,\ v\in \{0,1\}^n\right\} $$
\noindent where $Av\geq \mathbf{1}$ denotes being $\geq 1$ on all coordinates, what geometrically means being in a shifted cone.
\subsection{3-colorability as a spin glass}
As example of converting 3-colorability problem into e.g. global optimization problem, let us imagine 3 unit vectors of angles $0, 2\pi/3, 4\pi/3$ radians (0, 120, 240 degrees) corresponding to the three colors. Summing any two (different) of such vectors, we get the third one. Cosine of angle between them is $-1/2$.

Imagining there is some angle $\varphi_i$ assigned to each vertex, we get 3-coloring iff
$$\forall_{ij\in E}\ \cos(\varphi_i-\varphi_j)=-1/2$$
We can try Newton-Raphson method, or minimization of sum of $(\cos(\varphi_i-\varphi_j)+1/2)^2$ to stabilize in some relative angles (rotated by a constant value). From physical perspective we treat graph as a spin glass here, with nonstandard interaction: preferring 120 degree angles between neighboring spins.
\subsection{3-SAT as a combinatorial problem}
Imagine variables as $\{0,1\}$ values, and transform a 3-SAT formula into product of sums over clauses. Now summing over all possibilities, the formula is satisfied iff
$$0<\sum_{v\in\{0,1\}^n} (v_{11}+\bar{v}_{12}+v_{13})\cdot \ldots\cdot (\bar{v}_{m1}+v_{m2}+v_{m3})$$
\noindent where $v_{ij}$ is corresponding variable ($1\ldots n$), $\bar{v}=1-v$ corresponds to some specific negation pattern.

This sum can be decomposed into $3^m$ length $m$ monomials, however, for every monomial the summation is quite simple. If monomial contains both variable and its negation, the product is always zero. Otherwise, if monomial contains all $n$ variables, the summation gives $1$. If monomial contains only $n-k$ variables, the summation gives $2^k$. Hence, we have transformed 3-SAT into combinatorial problem of calculating numbers of monomials using given number of variables.

\section{Hamilton cycle problem through Grassmann numbers and differentiation}
In the Hamiltonian cycle problem, which is one of NP-complete, we want to determine if a given undirected graph has a cycle visiting all vertices exactly once. Denote by $A$ the $n\times n$ symmetric adjacency matrix of this graph. Powers of this matrix can be seen as sums over all paths, hence $\textrm{Tr}(A^n)$ can be seen as sum over all cycles.

However, this sum also contains paths going multiple times through a vertex - the problem is to "extract" Hamilton paths from this sum. Let us look at some two possibilities.
\subsection{Grassmann numbers}
Physicists working with fermionic fields use external algebra of Grassmann numbers $(\theta_i)$ which anti-commute~\cite{grass}:

$$\theta_i \theta_j = -\theta_j \theta_i$$

Hence, $\theta_i^2=0$ and having a sequence of such variables, we can sort them (changing sign) and such product vanishes if there are two or more identical terms there. Otherwise, the sign of such product is multiplied by the sign of applied permutation.

Observe that such Grassmann numbers seem a perfect tool for extracting Hamilton cycles from the $A^n$ approach. Denote diagonal $n\times n$ matrix with $n$ different Grassmann numbers on the diagonal as $D:=\textrm{diag}(\theta_i)$. Now diagonal terms of $(A\cdot D)^n$ are sums over all cycles, in which all those going twice through some vertex are vanishing due to anti-commutation. Hence it becomes sum of only Hamilton cycles.

Observe that such sum over all cycles can vanish due to cancellation of cycles of negative sign vertex permutations. One way to prevent that is using two different Grassman numbers for each vertex: $D':=\textrm{diag}(\theta_{2i}\theta_{2i+1})$. Thanks of it, permutation between two cycles will always need an even number of inversions - have $+1$ sign.

Finally, a given graph has a Hamilton cycle if and only if $(A\cdot D')^n$ has nonzero diagonal (or trace).\\

However, the issue with this approach is that Grassmann numbers are difficult to realize. A natural construction is using matrix of size $2^k$ for $k$ Grassmann numbers, for example for $k=2$:

$$\theta_1=\left(
             \begin{array}{cccc}
               0 & 0 & 0 & 0 \\
               1 & 0 & 0 & 0 \\
               0 & 0 & 0 & 0 \\
               0 & 0 & 1 & 0 \\
             \end{array}
           \right)
\qquad \theta_2=\left(
             \begin{array}{cccc}
               0 & 0 & 0 & 0 \\
               0 & 0 & 0 & 0 \\
               1 & 0 & 0 & 0 \\
               0 & -1 & 0 & 0 \\
             \end{array}
           \right)
$$

Finally, from the P$\ne$NP assumption we can conclude that indeed matrix representation of Grassmann numbers requires exponentially growing size of these matrices.

\subsection{Variable differentiation}
Another way to extract Hamilton paths from the $A^n$ approach is through introducing some variables (e.g. real or complex) and finally differentiating over them. Analogously as for Grassmann numbers, denote: $D=\textrm{diag}(x_i)$ as $n\times n$ diagonal matrix with $n$ different variables on the diagonal. Taking derivative over all these variables, we can extract the terms corresponding to Hamilton cycles from $(AD)^n$.

Beside just using $(AD)^n$, we could also for example use a function being a series of $AD$ and fix all the remaining powers to zero by the differentiation and final substitution of zeros to all variables. Finally, the existence of Hamilton cycle can be formulated as one of the three following equivalent conditions:

$$ 0\neq \frac{\partial^n}{\partial x_1\cdot\ldots\cdot\partial x_n} \textrm{Tr}\left((AD)^n\right)  $$
$$ 0\neq \frac{\partial^n}{\partial x_1\cdot\ldots\cdot\partial x_n} \textrm{Tr}\left(\exp(AD)\right)\Big|_{x_1=\ldots=x_n=0} $$
$$ 0\neq \frac{\partial^n}{\partial x_1\cdot\ldots\cdot\partial x_n} \textrm{Tr}\left((1-AD)^{-1}\right)\Big|_{x_1=\ldots=x_n=0} $$

\section{Geometric approaches for\\ graph isomorphism problem}
Graph isomorphism problem asks if given two indirected graphs are isomorphic, and is usually seen as essentially simpler than NP-complete, especially having in mind that in 2015 there was found quasi-polynomial algorithm~\cite{babai}.

Let us look at this problem from the perspective of linear algebra. For two graphs given by (symmetric) adjacency matrix $A$ and $B$, it is inexpensive to test if these matrices are similar by checking if their characteristic polynomials are identical: $\det(A-\lambda I)\equiv \det(B-\lambda I)$. In graph isomorphism problem we need to test if they are not only similar, but additionally if there exists similarity matrix which is permutation. While graph isomorphism problem restricts the matrices of interest here to $\{0,1\}$ adjacency matrices, we can expand this question to a general algebraic problem - if given two real or complex matrices differ only by permutation.

Positive test of their similarity means that they have identical eigenspectrum (with multiplicities). Hence a natural approach is trying to compare eigenvectors corresponding to the same eigenvalue - a nondegenerated eigenvector distinguishing some coordinates would bring restrictions to possible isomorphisms.

However, the real problem is for problematic cases: when eigenvalues are strongly degenerated, especially for (connected) Strongly Regular Graphs (SRG), which turn out to have only three different eigenvalues.

\begin{df} SRG($m,k,\nu,\mu$) contains $m$ vertex regular graphs: having $k$ degree for all vertices, and satisfying two additional conditions:
\begin{itemize}
  \item every two adjacent vertices have $\nu$ common neighbors,
  \item every two non-adjacent vertices have $\mu$ common neighbors.
\end{itemize}
\end{df}

They have to fulfill (\cite{cameron}): $(m-k-1)\mu=k(k-\nu-1)$ and their adjacency matrix $A$ fulfills two equations:
$$AJ=JA=kJ$$
$$A^2=k\mathbb{I}+\nu A+\mu(J-\mathbb{I}-A)$$
where $\forall_{ij}\, J_{ij}=1$ is $m\times m$ matrix built of all 1. It allows to conclude known analytic formulas for eigenstructure of strongly regular graphs: eigenvalue $k$ with multiplicity 1, and two ($\pm$) eigenvalues:
\be \lambda=\frac{1}{2}\left((\nu-\mu)\pm\sqrt{(\nu-\mu)^2+4(k-\mu)} \right)\label{eigval}\ee
with corresponding multiplicities:
\be \frac{1}{2}\left((m-1) \mp \frac{2k+(n-1)(\nu-\mu)}{\sqrt{(\nu-\mu)^2+4(k-\mu)}}\right).\label{eigmul}\ee

\begin{figure}[t!]
    \centering
        \includegraphics{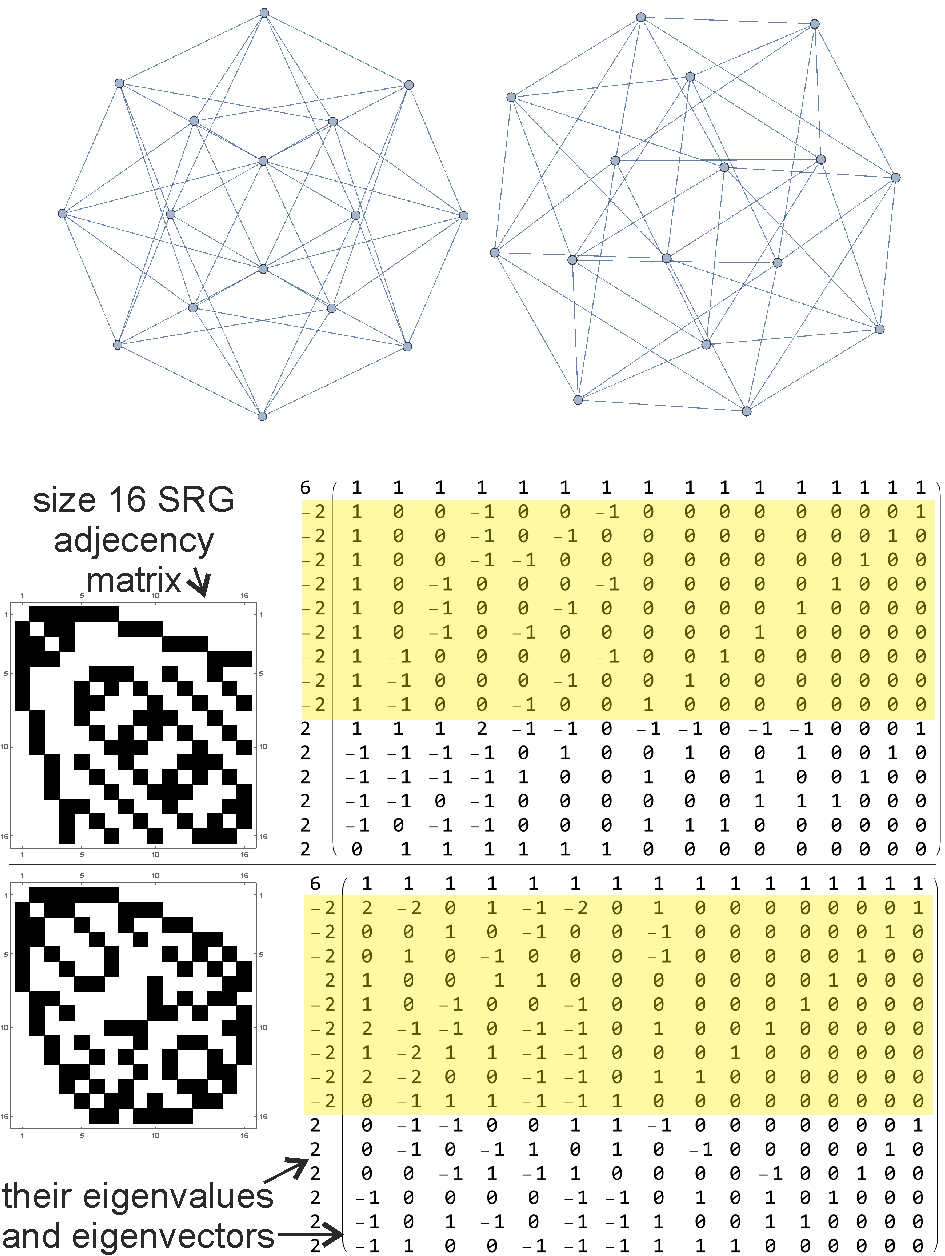}
        \caption{Example of two strongly regular graphs having the same parameters (16,6,2,2), which are not isomorphic. Eigenspaces of their adjacency matrices are strongly degenerated (1+9+6). The written eigenvectors are not orthogonalized. After Gram-Schmidt orthonormalization of such basis, and taking e.g. 16 "vertical vectors" of size 6 (or 9), we get very regular polyhedra in sphere in $\mathbb{R}^6$ (or $\mathbb{R}^9$): such that all neighbors correspond to vectors in a fixed angle below 90 degrees, all non-neighbors to another fixed angle above 90 degrees. The above example leads to two essentially different regular polyhedra: which cannot be rotated one into another.}
       \label{srg}
\end{figure}

Figure \ref{srg} contains two examples of SRG(16,6,2,2)\footnote{\url{http://www.maths.gla.ac.uk/~es/srgraphs.php}}. We see 1+9+6 eigenvalue degeneration. Let us focus on the last 6 rows: $6\times 16$ matrix representing 6 dimensional eigenspace, which can be analytically found (with terms being algebraic numbers) as kernel of $A-\lambda I$ using one of eigenvalues from (\ref{eigval}). The question is if we can get some information regarding potential isomorphism from such subspaces?

\subsection{Searching permutation among similarity matrices} \label{ortspace}
A direct approach is searching for permutation in the space of all possible similarity matrices (orthogonal) between $A$ and $B$. Assuming they are similar, we can diagonalize them to the same diagonal matrix $D$:
$$O_A A O_A^T = D= O_B^T B O_B \qquad B= (O_B O_A) A (O_B O_A)^T$$
hence $O_B O_A$ is a similarity matrix.

However, degenerated spectrum allows for $D=O D O^T$ for $O$ being orthogonal within each block of identical eigenvalues. Hence the space of similarity matrices between $A$ and $B$ can be characterized as:
\be \{P=O_B\, O\, O_A\ :\  D=O D O^T\} \ee
The graph isomorphism problem  asks if this set contains permutation - we could use a numerical procedure to try to find it there.

There are various ways to characterize permutation in the space of orthogonal matrices. For example defining $s_p=\sum_{ij} (P_{ij})^p$, orthogonality implies $s_2=n$. However, $s_3$ is usually below, it reaches $n$ only for permutation matrix.

Hence numerical approach might be looking for maximum of e.g. $\sum_{ij} ((O_B\, O\, O_A)_{ij})^3$, under $D=O D O^T$ condition, what can be obtained e.g. by interleaving gradient descent with Newton-Raphson step, or just searching in space of matrices being orthogonal within each eigenspace. However, such search might contain the general problem with optimization approaches: exponential number of local optimums, for example close to permutations not corresponding to the isomorphism.

\subsection{Geometrical interpretation of vertices}
Having $n\times m$ matrix representing a subspace - here $n=6$ vectors of length $m=16$, a natural first step is Gram-Schmidt ortonormalization, getting size $n$  orthonormal basis of this subspace. However, there is large freedom of choosing such othronormal basis: we can rotate it by multiplying such $n\times m$ matrix by any $n\times n$ orthogonal matrix.

There are invariants. Looking at vertical vectors in this $n\times m$ orthonormalized matrix: $x^i\in \mathbb{R}^n$, their norm does not change during  multiplication by $n\times n$ orthogonal matrix. Also $x^i \cdot x^j$ scalar products are rotation invariants. Grouping more such length $n$ vectors into matrix $M$, e.g. $\det(M^T M)$ (or trace) is also invariant.

\begin{figure}[t!]
    \centering
        \includegraphics{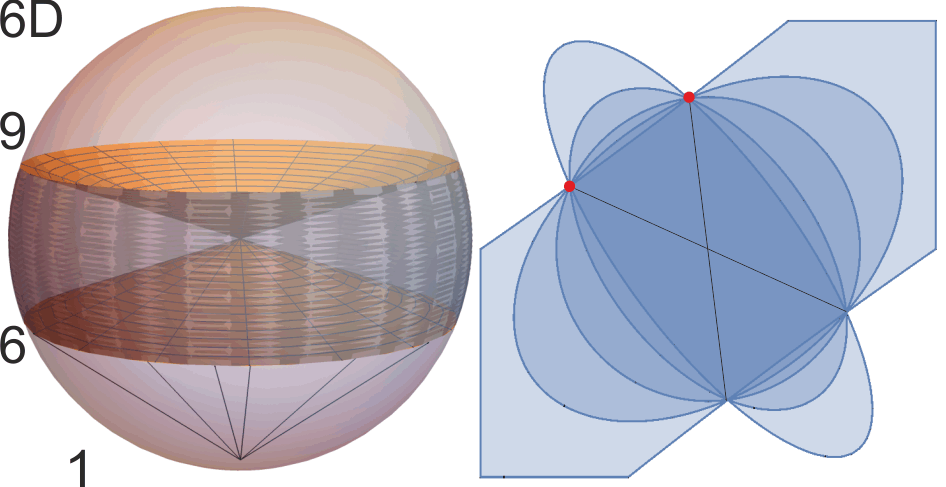}
        \caption{Left: schematic view of regular polyhedron formed of "vertical vectors" from eigenspace of adjacency matrix of strongly regular graph. For example 16 vertices in $\mathbb{R}^6$, which turn out to recreate relations from the graph: all 6 neighbors are in one fixed angle below 90 degrees, all the remaining are in another fixed angle above 90 degrees. Graph isomorphism test becomes question if two sets of points differ only by rotation - while locally they look similar, they should look essentially different for global rotation invariants. Right: describing our set of points $X$ (marked red) as intersection of ellipsoids/hyperboloids: $\bigcap_{P\in \mathcal{P}} \{x:x^TPx=1\}$ for $\mathcal{P}=\{P_0+a_1 P_1+\ldots+a_d P_d\}$ being $d$ dimensional affine space of symmetric matrices. It allows to use tools like rotation invariants to test differing by rotation. Such description adds symmetric points $X\to X\cup (-X)$, what is not a problem as it seems that in our case $X\cap (-X)=\emptyset$. A difficult question is if it does not add more points?}
       \label{polel}
\end{figure}

However, such invariants do not seem helpful for distinguishing SRGs, for example for single vectors (corresponding to individual vertices) their norms turn out identical - geometrically they are all on a single sphere centered in zero. Additionally, scalar product of such length $n$ vectors turns out to recreate the neighborhood relation of the graph: all neighbors are in one constant angle below 90 degrees, all non-neighbors are in another constant angle above 90 degrees:

\be x^i \cdot x^j = \left\{   \begin{array}{ll}                  \alpha \quad \textrm{if}\ i=j,\\
                  \beta \quad \textrm{if}\ i\neq j,\ A_{ij}=1,\\
                  \gamma \quad \textrm{if}\ i\neq j,\ A_{ij}=0.
                \end{array}
              \right.
\label{relation}
\ee

Where $\alpha,\ \beta, \gamma$ parameters can be expressed by parameters of SRG~\cite{cameron}. For simplicity let us normalize them to unit sphere: $x_i \to x_i/\sqrt{\alpha},\ \alpha\to 1,\ \beta\to \beta/\alpha,\ \gamma\to \gamma/\alpha$.

Hence, strongly regular graphs allow to construct very regular polyhedra on a sphere by taking orthonormal basis of eigenspace, then taking its "vertical vectors": $m$ vectors corresponding to individual vertices, of size $n$ being dimension of eigenspace. It provides an infinite discrete family of very nontrivial regular polyhedra, which might be useful to construct e.g. POVMs (positive-operator valued measure) or spherical designs.

\subsection{Testing directional orders}
Naive approach to search for graph isomorphism is testing all $m!$ vertex permutations. Geometrical interpretation obtained from looking at eigenspaces allows to emphasize rotation invariant subsets of orders. Specifically, choosing a unit vector (direction): $v\in\mathbb{R}^n$, we can sort the vertices accordingly to projection $x^i\cdot v$. Assume we focus only on vectors for which there are no two equal projections for our discrete set of points, what is a generic situation as we work on real numbers here.

\begin{figure}[t!]
    \centering
        \includegraphics{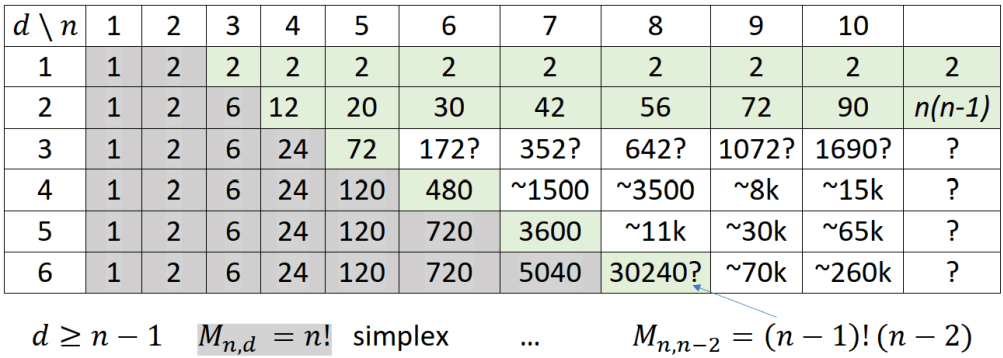}
        \caption{For $n$ points in $\mathbb{R}^d$, the question is maximal number of directional orders: accordingly to projection to some direction. In $d$ we have at most 2: left/right. For $n\leq d+1$ we can get any order. Generally the behavior is quite complex, the approximate values are from numerical search.}
       \label{orders}
\end{figure}

It is an interesting question\footnote{\url{https://math.stackexchange.com/questions/2599332/}} to find general boundary for the number of directional order for $m$ points in $\mathbb{R}^n$. However, while for low dimension $n$ the number of orders becomes essentially reduced while focusing only on directional, for $n\geq m-1$, we can get all $m!$ orders. Some numerically searched values are shown in Fig. \ref{orders}.

Hence in our case their number still grows exponentially. However, if we could construct a polynomial number of interesting directions (distinguishing all points), which have to agree for both sets, we could test only orders for such polynomial set of directions.

\subsection{Testing if two sets of points differ by rotation}
As discussed, looking at eigenspace we have converted the graph isomorphism test into question if two sets of $m$ points: $X$ and $Y$ in $\mathbb{R}^n$ (in Fig. \ref{srg}: $m=16,\ n=6$ or $9$) differ by rotation:
\be X \sim_r Y \quad \equiv\quad \exists_{O:O^TO=\mathbb{I}}\ X=\{Oy:y\in Y\}. \ee

Such sets turns out regular polyhedra, making a test based on local relations practically equivalent to the original problem. However, it gives the vertices additional geometric interpretation, and global structure which should be distinctive for non-isomorphic graphs. We can exploit it using (global) rotation invariants of such set. There are known rotation invariants: e.g. in 2D sum of squares of sine and cosine Fourier coefficients is rotationally invariant, in 3D we can use rotationally invariant (spherical) harmonics (discussed e.g. in \cite{shape}) - while they are difficult to generalize to higher dimensions, they (e.g. real spherical harmonics) base on homogeneous polynomials of coordinates, suggesting to use them here.\\

\subsubsection{Rotation invariants for homogeneous polynomials}
Treating rotation in $\mathbb{R}^n$ as $x\to Ox$ for orthogonal $O^T O=O O^T=\mathbb{I}$, there are well known complete sets of rotation invariants for degree 1 and 2 homogeneous polynomials - determining given polynomial up to rotation:
\begin{itemize}
  \item Degree 1 homogeneous polynomial: $p(x)=\sum_i p_i x_i$, after rotation $p_i\to \sum_a p_a O_{ai}$ has single invariant: $\sum_i p_i^2$, which characterizes it up to rotation (abstract class modulo rotation).
      $$\sum_i p_i^2=\sum_{ia\alpha} p_a O_{ai}\, p_\alpha O_{\alpha i}=\sum_{a\alpha}p_a p_\alpha \delta_{a\alpha}=\sum_a p_a^2 $$
  \item Degree 2 homogeneous polynomial: $p(x)=\sum_{ij}p_{ij} x_i x_j$ is just scaling in eigendirections of matrix of coefficients: $[p]:=[p_{ij}]_{ij=1\ldots n}$, set of $n$ eigenvalues (with multiplicities) determines this polynomial up to a rotation: we can use $n$ eigenvalues as a complete set of rotation invariants, or equivalently $\{\lambda^0,\ldots,\lambda^{n-1}\}$ coefficients of $\det([p]-\lambda \mathbb{I})$ characteristic polynomials, or equivalently $\textrm{Tr}([p]^\ell)=\sum_i \lambda_i^\ell$ for $\ell=1,\ldots,n$.
\end{itemize}
$\ $

For degree 3, homogeneous polynomial:\\ $p(x)=\sum_{ijk} p_{ijk} x_i x_j x_k$ is transformed by rotation using $p_{ijk}\to\sum_{abc} p_{abc} O_{ai} O_{bj} O_{ck}$. We can check that for example $\sum_{ijk}p_{ijk}p_{jki}=$
$$=\sum_{ijk}\left(\sum_{abc}p_{abc}O_{ai}O_{bj}O_{ck}\right)
\left(\sum_{\alpha\beta\gamma} p_{\beta\gamma\alpha} O_{\beta j}O_{\gamma k} O_{\alpha i}\right)= $$
\noindent $=\sum_{abc} p_{abc}p_{bca}$ is rotation invariant using the $\sum_i O_{ai} O_{\alpha i}=\delta_{a\alpha}$ relation for $ijk$ indexes.

\begin{figure}[t!]
    \centering
        \includegraphics{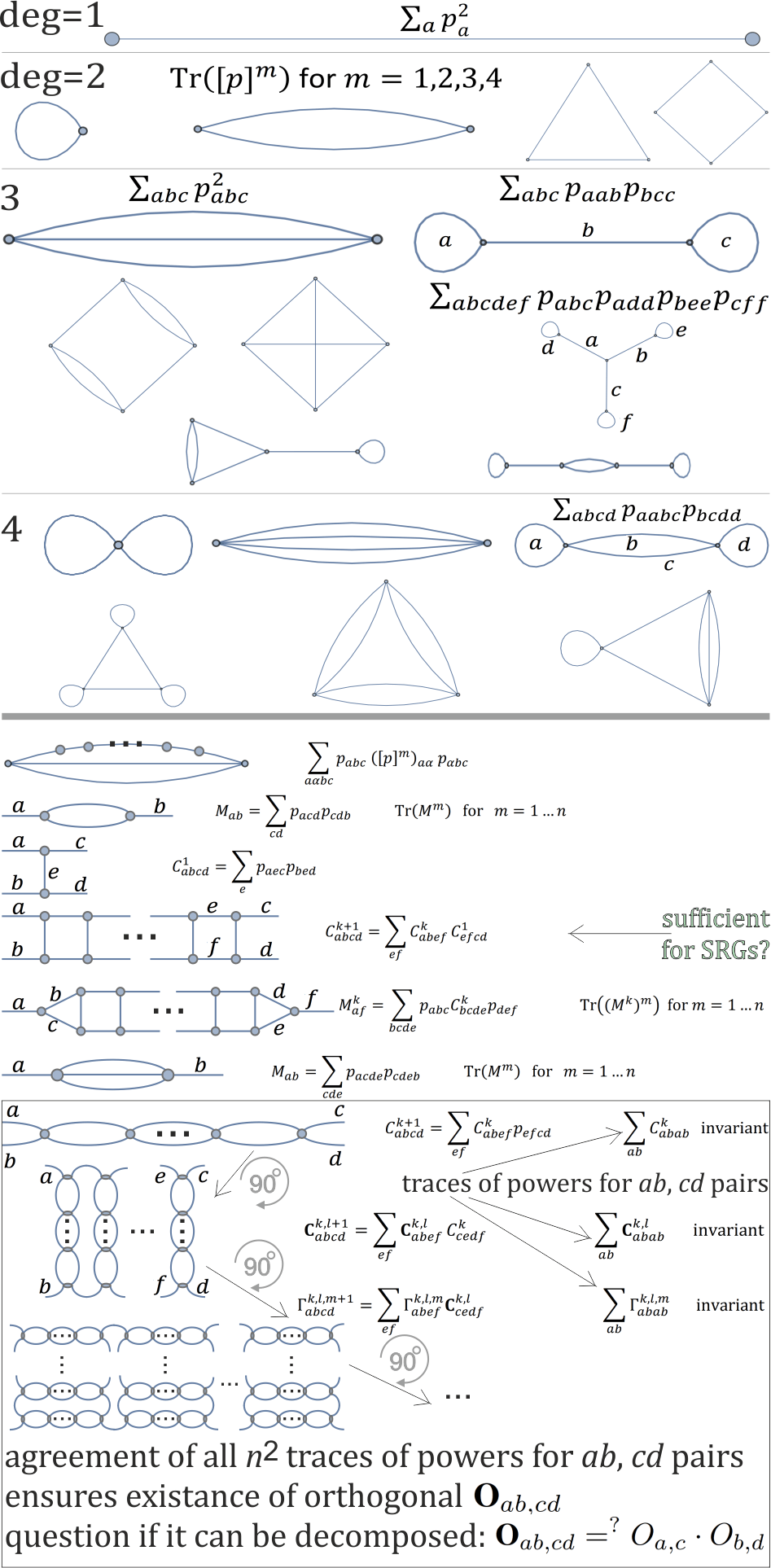}
        \caption{Top: examples of diagrammatic representations of rotation invariants for degree 1, 2, 3, 4 homogeneous polynomials. Each vertex corresponds to term of polynomial and has the same degree as polynomial. Operating on commutative fields like $\mathbb{R}$ here, edges for given vertex are indistinguishable. Every edge corresponds to summation over corresponding index, like in matrix product, and is rotation invariant thanks to $\sum_i O_{ai}O_{\alpha i}=\delta_{a\alpha}$. Invariants from disconnected graphs can be omitted as being products over its components. Bottom: examples of systematic construction of large number of rotation invariants, especially the framed family. They use traces of powers for pairs of variables (tensor product), which agreement on all $n^2$ powers allows to conclude existence of orthogonal matrix for pairs: $\textbf{O}_{ab,cd}$. We are interested in differing by rotation of single variables, what becomes the question of decomposition: $\textbf{O}_{ab,cd}=^? O_{a,c}\cdot O_{b,d}$. We can analogously construct higher order invariants by rotating previous diagram 90 degrees and combining multiple copies, getting practical formula like written for $\Gamma$, or analogously for higher orders. For higher degree polynomials we can increase the numbers of joining edges, e.g. from 2+2 for degree 4 to 3+3 or 4+2 for degree 6.}
       \label{diag}
\end{figure}

Analogously we can construct such invariants by using exactly two copies of indexes we sum over, allowing for their diagrammatic representation - some examples are presented in Fig. \ref{diag}. Working on commutative field like $\mathbb{R}$, we can assume that $p$ is identical for all permutation of indexes, e.g. $[p]=[p]^T$ for degree 2. Hence both vertices and edges are indistinguishable - every such graph corresponds to a single rotation invariant.

However, some of them might be dependent, e.g. $\sum_{ab} p_{aa}p_{bb}=\left(\sum_a p_{aa}\right)\left(\sum_b p_{bb}\right)$, what would be represented by disconnected graph - hence it is sufficient to focus on connected graphs.

There are also more sophisticated dependencies, e.g. $\textrm{Tr}([p]^{n+1})$ can be calculated from $\textrm{Tr}([p]^{\ell})$ for $\ell=1,\ldots,n$. This is caused by the fact that degree 2 homogeneous polynomial is defined by eigenvalues ($\Lambda$) and rotation ($O$): $[p]=O^T \Lambda O$. Hence rotation invariant are exactly $n$ parameters of diagonal $\Lambda$, which often are uniquely determined by $n$ independent algebraic equations.

However, for degree 3 and higher the situation seems much more complex. The number of independent parameters we can optimize with $O(n)$ matrix, like in $[p]=O^T \Lambda O$ case, is $n(n-1)/2$. The number of parameters in symmetric matrix is $n(n+1)/2$, hence optimization over orthogonal matrices allows to reduce the number of independent parameters to the difference: $n$, what agrees with matrix $\Lambda$. Degree $g$ homogeneous polynomial analogously have ${{n+g-1}\choose g}$ parameters, so after optimization with orthogonal matrix there still remains $O(n^g)$ parameters.

This approach can be also applied for polynomials which are not homogeneous~\cite{rotinv}, by just analogously using graphs with vertices of varying orders. Such mixed terms describe relative rotation between homogeneous parts, which contains at most $n(n-1)/2$ parameters.

Such rotational invariants could be useful also e.g. in machine learning problems like image recognition: describe patterns using homogeneous polynomials, calculate their rotation invariants, and compare invariants to test (screen for) similarity with objects of an unknown rotation.

\subsubsection{Describing a set as intersection of ellipsoids}
In our case, set $X$ is e.g. 16 point in $\mathbb{R}^6$, hence degree 1 homogeneous polynomial does not have sufficient number of parameters to agree in all points. However, degree 2 is sufficient so we will focus on it. If required, higher degrees homogeneous polynomials might be also considered.

A standard way for describing a set with 2nd order polynomial is using covariance matrix in Principal Component Analysis method, however, here we need to uniquely determine exactly a given set of points $X$. For this purpose, we can for example look at polynomials with fixed values on all $m$ points of $X=\{x^k\}$:
\be \forall_k\ \sum_i P_{ii} \left(x_i^k\right)^2+ 2 \sum_{i<j} P_{ij} x_i^k x_j^k = w_k \label{leq} \ee

The number of $\{P_{ij}\}$ coefficients ($P=P^T$) is $D=n(n+1)/2$. Hence, having $|X|=D$ points, and assuming that this system of linear equation is well defined ($\det\neq 0$), we would get a unique polynomial $p(x)=\sum_{ij} P_{ij} x_i x_j$ agreeing on $X$, for which differing by rotation can be tested e.g. using characteristic polynomial.

As we want the points to be indistinguishable, let us choose all $w_k=1$. This way $P$ defines ellipsoid $\{x:x^TPx=1\}$, which contains much more points than discrete set $X$ we would like to describe. However, if the number of points is smaller: $|X|=D-d$, the set of linear equations (\ref{leq}) defines \textbf{affine space of symmetric matrices}:
\be \mathcal{P}=\{P_0+a_1 P_1+\ldots a_d P_d:\ a\in\mathbb{R}^d\} \ee
\noindent and analogously $\mathcal{Q}$ for the second set $Y$. Such $\mathcal{P}$ defines our points as intersection of ellipsoids/hyperboloids: $\{x:\ \forall_{P\in \mathcal{P}}\ \ x^TPx=1\}$ like in Fig. \ref{polel}. Ths way the original question if our sets differ by rotation $(X \sim_r Y)$ becomes question if
\be\mathcal{P}\sim_s\mathcal{Q}\quad\equiv\quad\exists_{O:O^TO=\mathbb{I}}\ \mathcal{P}=\{O^T QO:Q\in\mathcal{Q}\}. \label{sims}\ee

Testing if two sets differ only by rotation is the most difficult when they are on a sphere, what turns out true in our case, and allows to assume that $P_0=Q_0\propto \mathbb{I}$ and remove it from $\mathcal{P}$ and $\mathcal{Q}$ while testing if $\mathcal{P}\sim_s\mathcal{Q}$.

We know that $x^k$ points are in $\{x:x^TPx=1\}$ ellipsoids/hyperboloids for all $P\in\mathcal{P}$. However, all these ellipsoids are symmetric: contain also $-x^k$. Hence we have
\be X\cup(-X)\ \subset\ \{x:\ \forall_{P\in \mathcal{P}}\ \ x^TPx=1\} \label{hyp}\ee
The fact that this description adds symmetric points seems not an issue here as in our case the set seem not to contain symmetric points: $X\cap (-X)=\emptyset$ (yet to be proven). The question is if other points are not added in this description as intersection of ellipsoids? Or more formally: if we can conclude $\mathcal{P}\sim_s \mathcal{Q}\ \Rightarrow\ X\sim_r Y$ ?

Using less than $D$ points as constraints, we could take some other point and enforce any value there - hence, such point will not be in the intersection. However, this argument requires linear independence in (\ref{leq}) for this new point, what turns out a very complicated condition. Using higher dimension ($d$) provides stronger restriction, making it more likely to give sufficient condition, but it is yet to be proven.\\

To practically calculate $\mathcal{P}$, let us transform symmetric matrices $P$ into length $D$ vectors in the following way:
$$V(P):=(P_{11},\ldots,P_{nn},\sqrt{2}P_{12},\ldots,\sqrt{2} P_{n-1,n}).$$
Thanks of multiplying the nondiagonal terms by $\sqrt{2}$, natural scalar product on these vectors is recreating Frobenius inner product $\langle \cdot,\cdot \rangle_F$ of these matrices:
$$\langle P,Q\rangle_F:=\textrm{Tr}(PQ^T)=\sum_{ij} P_{ij} Q_{ij}= V(P)\cdot V(Q)$$
which is invariant while rotation $P\to O^TPO,\ Q\to O^TQO$ and induces Frobenius norm: $\|P\|_F^2=\textrm{Tr}(PP^T)=\sum_{ij} P_{ij}^2$.

Using rotation invariant inner product, we get geometry in $\mathcal{P}$ which has to be identical as in $\mathcal{Q}$ if indeed $\mathcal{P}\sim_s \mathcal{Q}$. Orthonormal basis in $\mathcal{P}$ has to be also orthonormal in $\mathcal{Q}$, different can differ only by rotation inside $\mathcal{Q}$.

Now the (\ref{leq}) can be written as $V(P)\cdot v(x^k)=w_k$ for:
$$v(x):=(x^2_1,\ldots,x_n^2,\sqrt{2} x_1 x_2,\ldots,\sqrt{2}x_{n-1} x_n).$$

Finally, to construct orthonormal basis of $\mathcal{P}$ (and analogously $\mathcal{Q}$), we can take $|X|=D-d$ vectors $v(x^k)$, append $D$ canonical vectors at the end: $(\delta_{ij})$ for $j=1,\ldots,D$. Then perform Gram-Schmidt othogonalization for these $|X|+D$ vectors. Now the nonzero vectors in the appended positions create orthonormal basis of $\mathcal{P}$.\\

Having such orthonormal bases, $\mathcal{P}\sim_s \mathcal{Q}$ iff
\be \det(a_1 P_1+\ldots+a_d P_d - \lambda \mathbb{I})=\det(b_1 Q_1+\ldots+b_d Q_d - \lambda \mathbb{I}) \label{char}\ee
for $b_i=\sum_j U_{ij} a_j$ and some  $U^T U=U U^T=\mathbb{I}$. Observe that rotation $U$ is different than in (\ref{sims}) definition of $\sim_s$.

A direct approach for such test it taking successive $P_j$ and checking if $\{b_1 Q_1+\ldots+b_d Q_d\}$ contains matrix of the same eigenspectrum as $P_j$, what seems achievable using numerical methods, e.g. by interleaving steps of QR algorithm and optimization of $b_i$ coefficients. A possible difficulty here is large number of such solutions, which could make it difficult to ensure the required condition that all $P_j$ correspond to the same rotation $O$ in (\ref{sims}).\\

\paragraph{Approach and remarks}
Generalized characteristic polynomial (\ref{char}) can be directly calculated for $d=1$, however, we need to work on larger $d$ for certainty of not introducing additional points in such description, and to work on the actual number of points $|X|=m$. If required, the number of points $X$ can be altered, for example by using points in the centers of all edges instead - which allow to recreate original vertices by averaging over neighbors and projecting to sphere.

An example of practical approach to directly work on larger $d$ is looking at $\textrm{Tr}(P^\ell)$ for $\ell=3,4,\ldots$, which turns out homogeneous polynomials, e.g.
\be \textrm{Tr}(P^3)=\sum_{ijk} \textrm{Tr}(P_{i} P_{j} P_{k}) a_i a_j a_k \ee
Differing by rotation can be tested for homogeneous polynomials using rotation invariants discussed earlier - every graph for degree 3 and 4 in Fig. \ref{diag} defines one rotation invariant we could use here. Being equal for corresponding invariants of $\mathcal{P}$ and $\mathcal{Q}$ is a necessary condition for graph isomorphism.

This way the final algorithm is: for strongly regular graph given by $m\times m$ adjacency matrix $A$, calculation of proposed invariants consists of the following steps:
\begin{enumerate}
  \item Choose and calculate eigenvalue $\lambda$ and its multiplicity $n$ using formulas (\ref{eigval}) and (\ref{eigmul}),
  \item Calculate orthonormal basis of kernel of $A-\lambda \mathbb{I}$, denote columns of such $n\times m$ matrix as $X=\{x^k\}_{k=1\ldots m}$ set of points, where $x^k\in\mathbb{R}^n$,
  \item Calculate $P_1,\ldots,P_d$ orthonormal basis of $\mathcal{P}$,
  \item Calculate terms of homogeneous polynomials like $p_{ijk}=\textrm{Tr}(P_{i} P_{j} P_{k})$,
  \item Calculate rotation invariants using formulas like in Fig. \ref{diag}.
\end{enumerate}
However, getting a complete set of invariant: which equality would allow to imply $\mathcal{P}\sim_m \mathcal{Q}$, is a complicated problem, e.g. might require to use $\textrm{Tr}(P^\ell)$ for relatively large $\ell$.

Finally there are two missing steps to complete such graph isomorphism test:
\begin{enumerate}
  \item Show that for interesting sets $X$ (e.g. fulfilling $X\cap (-X)=\emptyset$) and sufficiently large dimension $d$, described intersection of ellipsoids does not add too many artificial points:
  $$\mathcal{P}\sim_s \mathcal{Q}\quad \Rightarrow\quad X\sim_r Y $$
  \item Efficient testing of $\mathcal{P}\sim_s \mathcal{Q}$, e.g. through equality of sufficiently large set of invariants.
\end{enumerate}

\subsection{Complete description using degree 6 polynomial}
Let us take another look at testing graph isomorphism in geometric interpretation: each of $m$ vertices gets $x^i\in\mathbb{R}^n$ point, they form a regular polyhedron ($\|x^i\|=1$, $x^i\cdot x^j=\beta$ for neighbors, $x^i\cdot x^j=\gamma$ for non-neighbors (\ref{relation})), and we need to test if two sets ($X$ and $Y$) satisfying this relation differ only by rotation.

While the previously discussed description was based on low degree: quadratic polynomial, one of its remaining questions is certainty if/when it does not add too many artificial points. In contrast, a certain complete description of such discrete set can be easily made with higher degree polynomials, e.g. $p(x)=\prod_i \|x-x^i\|^2$ or lower degree: $\prod_i (x\cdot x^i-1)$ using additional constraint of being on unit sphere $(\|x\|=1)$. However, these are $O(m)$ degree polynomials, making that test ensuring that two such polynomials differ only by rotation would rather require exponential cost.

Fortunately, in SRG case these sets of points fulfill simple relation (\ref{relation}): that scalar product between such two vectors can only have one of three values: $1,\ \beta$ or $\gamma$. It allows to use degree 6 polynomial for description as intersection of unit sphere and $m$ triples of hyperplanes:
\be p(x) =\sum_{i=1}^m (x\cdot x^i -1)^2 (x\cdot x^i -\beta)^2 (x\cdot x^i -\gamma)^2 \label{deg6} \ee
The first question is if it forms a complete description:

$$ X =^? \{x:p(x)=0,\ \|x\|=1\} $$
The "$\subset$" part comes from (\ref{relation}). However, "$\supset$" is not generally true and requires to use the fact that the number of points ($m$) is lager than dimension ($n$), as intersection of $n$ linearly independent hyperplanes determines unique point in $\mathbb{R}^n$. In our case, choosing the smaller nontrivial eignespace (\ref{eigmul}), we have certainty that $m>2n$. Hence, artificially adding new points seem unlikely for such description, but it still would require a formal proof.

However, in our case this proof turns out unnecessary as we know the numbers of vertices for both our sets: that each set contains 1 vertex with $x\cdot x_i=1$, (degree) $k$ vertices $x\cdot x_i=\beta$ and $m-k-1$ vertices with $x\cdot x_i=\gamma$. This additional constraint for the number of points inside the three hyperplanes forbids to use more points than already certain $X$, making our degree 6 polynomial $p(x)$ a complete description of $X$:
\be X=\{Oy:y\in Y\}\quad \Leftrightarrow\quad p(x)=q(Oy)\ee
where $O^T O=OO^T=\mathbb{I}$ and $q$ is analogous degree 6 polynomial (\ref{deg6}) for $Y$ set describing the second graph.\\

Finally we have converted the problem if two strongly regular graphs are isomorphic, into a problem if two degree 6 polynomials $p$ and $q$ differ only by rotation. Figure \ref{diag} discusses how to construct rotation invariants for polynomials, more discussion can be found in \cite{rotinv}.\\

\subsubsection{Successful distinguishing with degree 3 polynomial}
In $\mathbb{R}^n$ we have $m\approx 2n$ vectors of fixed length, hence the number of degrees of freedom of our problem is $\approx 2n^2$. Additionally, we are not interested in their absolute rotation, allowing to subtract $\approx n^2/2$ degrees of freedom.

In contrast, above description uses degree 6 polynomials - having in general $\approx n^6/6!$ degrees of freedom. It suggests that we can use lower degree polynomial for complete description. While this still requires formal proof, tests suggest that homogeneous degree 3 polynomial $(p(x)=\sum_{i=1}^m \left(x_i\cdot x\right)^3)$ might be sufficient: successfully distinguish strongly regular graphs up to degree 29: by testing equality of eignspectrum (through traces of powers) of $n^2\times n^2$ matrix on pairs (tensor product):
\be M_{ab,cd}=\sum_{i,j=1}^m x^i_a \, x^i_c\, (x^i\cdot x^j)\, x^j_b \, x^j_d \label{invform} \ee
Its traces of powers $(\textrm{Tr}(M^k))_{k=1..n^2}$ are invariants corresponding to marked ladder-like graph in Fig. \ref{diag}, and agreement of all $n^2$ for both graphs ensures existence of orthogonal $\textbf{O}_{ab,cd}$ - shifting the main question to possibility of its decomposition: $\textbf{O}_{ab,cd}=^? O_{a,c}\cdot O_{b,d}$.

The discussed two 16 vertex SRGs have equal trace of first and second power of this matrix, however, the third one: $\textrm{Tr}(M^3)$ turns out different for two graphs. Further case of multiple strongly regular graphs with the same parameters are 15 graphs for 25 vertices, which turned out distinguished by nearly all traces of power, including the first one: $\textrm{Tr}(M)$. It turned also true for the next three cases: 10 graphs with 25 vertices, 4 graphs with 28 vertices and 41 graphs with 29 vertices. The next available case are 3854 graphs for 35 vertices (not tested yet).\\

Obviously this is not a proof that this polynomial time algorithm (directly $O(n^6)$) will distinguish all strongly regular graphs with its $n^2$ invariants. If so: there are found two strongly regular graphs with identical all $n^2$ invariants, we can analogously use other invariants with constructions like in Fig. \ref{diag}, using higher degree polynomial, for example:
$$ M_{ab,cd}=\sum_{i,j=1}^m x^i_a \, x^i_c\, (x^i\cdot x^j)^s\, x^j_b \, x^j_d\quad \textrm{for any power } s$$
$$ M_{ab,cd}=\sum_{i,j,k=1}^m x^i_a \, x^i_c \, (x^i\cdot x^j)^s \, (x^j\cdot x^k)^s \, x^k_b \, x^k_d $$
$$ M_{abc,def}=\sum_{i,j=1}^m x^i_a \, x^i_b \, x^i_d \, (x^i\cdot x^j)^s \, x^j_c \, x^j_e \, x^j_f$$
$$M_{abc,def}=\sum_{i,j,k=1}^m x^i_a  x^i_d x^j_b x^j_e x^k_c x^k_f (x^i\cdot x^j)^s(x^i\cdot x^k)^s(x^j\cdot x^k)^s$$

For formal proof we need to show that some set of invariants is complete: determines modulo rotation. It can be done directly for points (using restriction for their number e.g. $m\leq 2n+1$ ), or for polynomials of degree 6 or lower (down to 3) - if shown that it provides complete description.\\

\subsubsection{Constructions restricting rotations to permutations for general graph}
While the previous considerations were for strongly regular graphs (number of vectors $\approx 2\cdot$ dimension), the question if two graphs are isomorphic can be translated into question if two sets of vectors differ only by rotation also for general graphs ($|V|+|E|$ vectors in $\mathbb{R}^{|V|}$ for construction below).

Specifically, for graphs with $n$ vertices $V$ and edges $E$, take set of vectors: $X=\{e_1,\ldots,e_n\}\cup\{e_i+e_j:(v_i,v_j)\in E\}$ for canonical basis $\{e_i\}$.

This set contains $n$ vectors of length 1 forming orthogonal basis, and the remaining have length $\sqrt{2}$. Constructing such sets $X,Y$ for two graphs, rotation between them $X=OY$ needs to maintain the length 1 vectors, hence $O$ needs to be a permutation by this choice of points.

We can describe this set of points e.g. by $n^2\times n^2$ matrix $M_{ab,cd}$ constructed using formula (\ref{invform}), then test graph isomorphism by checking similarity of such matrices constructed for both graphs - it is necessary condition, the question is how to make it sufficient with a proof?

For this construction, matrix $M_{ab,cd}$ uniquely determines the original adjacency matrix: for example by looking at $M_{aa,cc}$, which is nonzero iff $ac$ is edge or $a=c$.


Hence, we have $M_{ab,cd}$ constructed for one graph, and let say $\bar{M}_{ab,cd}$ analogously constructed for the second graph - each one uniquely determining the corresponding graph. Testing $\textrm{Tr}(M^\ell)=\textrm{Tr}(\bar{M}^\ell)$ for $\ell=1\ldots n^2$, what can be done in polynomial time including construction, we know that they are similar: there exists orthogonal $n^2\times n^2$ matrix $\mathbf{O}$ such that $\mathbf{O}M\mathbf{O}^T=\bar{M}$.

To prove sufficiency of this test, there has remained to show that $\mathbf{O}$ can be decomposed $\mathbf{O}_{ab,cd}=O_{a,c}\cdot O_{b,d}$ for some orthogonal $O$, which have to be a permutation due to construction of set of vectors.

Another approach might be using the fact that both $M$ and $\bar{M}$ have coefficients being small natural numbers (nonzero only for near vertices) for the discussed construction from graphs. It makes it unlikely to satisfy $\mathbf{O}M\mathbf{O}^T=\bar{M}$ for $\mathbf{O}$ with coefficients not being small natural numbers, what combined with orthogonality implies being a permutation.

We can also contract such orthogonal matrix on tensor product: e.g. define $O_{a,c}=n^{-1/2} \sum_{bd} \mathbf{O}_{ab,cd}$, but it is not necessarily orthogonal. Neither is $O_{a,c}=\mathbf{O}_{aa,cc}$, but this kind of gluing of graph vertices leads to experimentally successful test discussed in the next subsection.

While this approach might be less practical than discussed in the next section, it seems more promising for getting a formal proof for sufficient condition in polynomial time. We can analogously use different constructions of e.g. $n^2\times n^2$ matrices, which determine the original adjacency matrix and enforce permutation by construction. For example constructions like $M=\sum_{v\in W} ww^T$ for $W$  using e.g. $e_{ii}+e_{ij}+e_{ji}+e_{jj}$ canonical $n^2$ dimensional vectors for all $ij$ edges, to which we need to add some "gadgets" enforcing permutation.

\subsection{Experimentally successful deformation-based invariants}
Having two vectors $x,y\in\mathbb{R}^n$, equality of their Euclidean norm: $\sum_i x_i^2=\sum_i y_i^2$ leaves a continuous space of possible rotations between them. If we additionally test belonging to the same e.g. 4-th norm ball: $\sum_i x_i^4=\sum_i y_i^4$, the space of possible rotations becomes much smaller, e.g. discrete for $n=2$. It can be imagined as using $x_i \to x_i^2$ deformation on both vectors before testing if they are on the same Euclidean sphere. This way we can get certainty that they differ only by signs and permutation of coordinates if testing for all $\ell=1..n$ deformations $x_i \to x_i^\ell$, as symmetric polynomials determine sets of values (without order, requires number of polynomials being size of set).

Let us try to take this intuition to restrict $\mathcal{O}=\{O:O^TO=\mathbf{I},\ O^TAO=B\}$ space of possible rotations between $A,B$ similar (adjacency) matrices, to permutations only - by testing similarity not only for the original matrices, but also for their deformed versions: test
\be\forall_{t\in T}\,\forall_{\ell=1..n}\ \textrm{Tr}(t(A)^\ell)=\textrm{Tr}(t(B)^\ell)\ee for some finite  set of deformation $T$, containing identity. If a polynomial number of such tests is sufficient to restrict $\mathcal{O}$ to permutations only, we would get polynomial test for (general) graph isomorphism problem this way.

The question is how to choose such matrix deformations? Standard matrix functions $f(A)$ do not allow for it: if $A$ and $B$ are similar then $f(A)$ and $f(B)$ are also similar.

For standard matrix functions we can use Taylor expansion, expressing this matrix function with matrix polynomials. Like in Fig. \ref{diag}, we can represent $A^k$ using line graph: with $k-1$ degree 2 vertices representing summation over intermediate indexes - corresponding to standard matrix multiplications.

Let us generalize this construction to defined by any graph: not restricted to degree 2, with kind of multiplication of multiple matrices, e.g. for (degree) three: $\sum_d A_{ad} A_{bd} A_{cd}$. Using invariants as traces of powers of matrices deformed this way, they remain permutation invariant - as permutation would only change order of summation. However, it turns out that they are not necessarily invariants for other, orthogonal but not permutation, changes of basis. For example here is a generalized matrix function coming from gluing two vertices of (\ref{invform}), or directly from "$\triangleleft\hspace{-0.3mm}\triangleright$"-like graph with degree 3 vertices:

\be t(A)_{ab}=\sum_{ij} A_{ai} A_{aj} A_{ij} A_{ib} A_{jb} \label{tform}\ee

Tested simpler graphs-based generalized matrix functions were not successful for distinguishing SRGs here, but this one was. Additionally, for $A,B$ being two 16 vertex SRGs discussed earlier, while both have the same eigenspectrum with 1+6+9 degeneracy, $t(A+\mathbf{I})$ has the same degeneracy, but $t(B+\mathbf{I})$ has different: 1+6+6+3. So this strange generalized matrix function was able to split 9 dimensional eigenspace into 3+6 subspaces, what is a quite surprising property.

For a larger test: 41 SRGs of 29 vertices and the same spectrum, $\textrm{Tr}(t(A)),\textrm{Tr}(t(A)^2),\textrm{Tr}(t(A)^3),\textrm{Tr}(t(A)^4)$ are presented below, calculated in $O(|V|^4)$ time:
\begin{center}
\noindent$2436, 394632, 138735072, 65369475360$\\
$2436, 400536, 139953984, 65542520928$\\
$2436, 400536, 139960416, 65544296160$\\
$2436, 400680, 139981824, 65546337952$\\
$2436, 400680, 139992192, 65549199520$\\
 $2436, 401688, 140192640, 65576335200$\\
 $2436, 401688, 140198688, 65578004448$\\
 $2436, 401720, 140200848, 65578282848$\\
 $2436, 401720, 140203728, 65579077728$\\
 $2436, 401784, 140212800, 65579292000$\\
 $2436, 401784, 140212992, 65579874912$\\
 $2436, 401784, 140214960, 65580457056$\\
 $2436, 401784, 140216112, 65580775008$\\
 $2436, 401784, 140218080, 65581279200$\\
 $2436, 401784, 140218272, 65580802272$\\
 $2436, 401816, 140217744, 65580344544$\\
 $2436, 401816, 140220432, 65581141216$\\
 $2436, 401816, 140221536, 65581382496$\\
 $2436, 401816, 140222016, 65581527648$\\
 $2436, 401816, 140222304, 65581607136$\\
 $2436, 401816, 140222784, 65581726944$\\
 $2436, 401816, 140223888, 65582095072$\\
 $2436, 401816, 140226576, 65582782176$\\
 $2436, 401848, 140225424, 65581545568$\\
 $2436, 401848, 140226336, 65581872480$\\
 $2436, 401848, 140231232, 65583223776$\\
 $2436, 401848, 140232144, 65583400288$\\
 $2436, 401880, 140233776, 65583047520$\\
 $2436, 401880, 140237040, 65583948384$\\
 $2436, 401912, 140237280, 65583099744$\\
 $2436, 401912, 140241936, 65584683872$\\
 $2436, 401912, 140242128, 65584736864$\\
 $2436, 401912, 140246784, 65585722848$\\
 $2436, 401944, 140246064, 65584881248$\\
 $2436, 401944, 140251248, 65586312032$\\
 $2436, 401976, 140251392, 65585296480$\\
 $2436, 401976, 140253744, 65586095328$\\
 $2436, 401976, 140256816, 65586943200$\\
 $2436, 401976, 140259168, 65587442656$\\
 $2436, 402024, 140262192, 65586962592$\\
 $2436, 402024, 140268240, 65588631840$\\
\end{center}
We see that the first power gives the same value ($2436$) for all - do not distinguish any of these graphs. Second power has some distinguishing power, but trace of third or fourth power alone turns out unique for all 41 graphs. Sorting them lexicographically like above, we get a linear order, for example among SRGs of given parameters.

\begin{figure}[t!]
    \centering
        \includegraphics{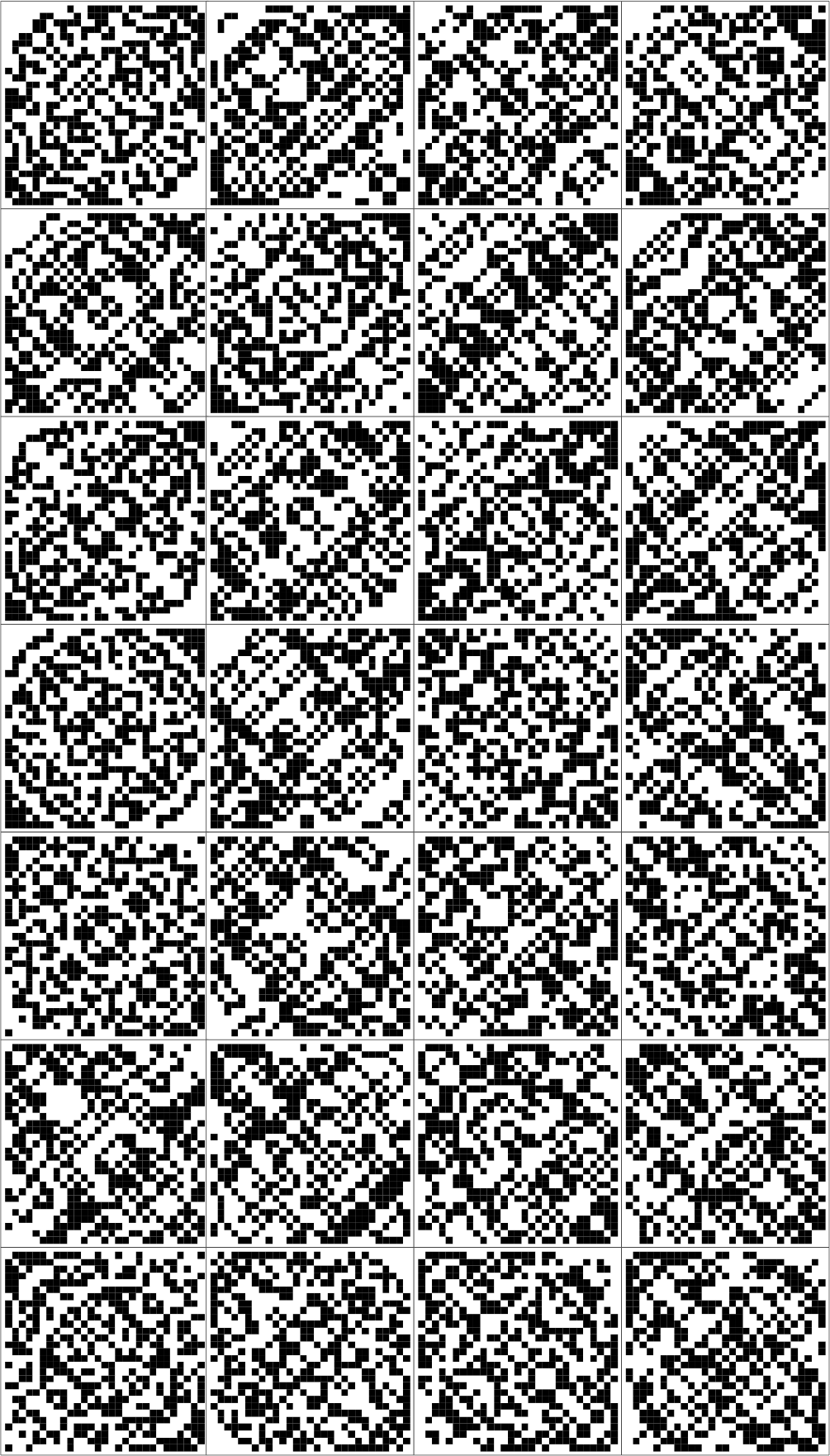}
        \caption{Adjacency matrix $A$ of the same 29 vertex SRG using 28 different orders of vertices defined by eigenvectors. While spectrum of $A$ has only 3 different eigenvalues, $t(A)$ has maximal: 29 unique eigenvalues - $t$ has completely removed initial strong degeneracy. Its first eigenvalue has equal coordinates: do not allow to define order among vertices, but the remaining 28 have nearly unique coordinates, allowing to define vertex order in nearly unique way - getting presented 28 adjacency matrices using vertex order from 2nd, ... , 29th eigenvector.}
       \label{29ord}
\end{figure}

The number of distinct eigenvalues of $t(A)$ turns out to take one of three possibilities here: 3, 21, or 29 - in the last case it removes all degeneracy, coordinates of such non-degenerated eigenvectors allow e.g. to nearly uniquely order vertices in 28 ways - presented in Fig. \ref{29ord} for one of these graphs.

Possibility to uniquely order vertices implies that this graph has no nontrivial automorphisms. From the other side, large automorphism group would prevent reducing degeneracy by such deformations and their permutation invariants, suggesting e.g. that the $t(A)$ case with only 3 unique eigenvalues has very large automorphism group. Using presented method to find automorphism group is an interesting perspective for further work.

To summarize, the following condition is necessary and usually sufficient to test if $A,B$ $n\times n$ matrices differ only by permutation:
\be \forall_{\ell=1..n} \textrm{Tr}(A^\ell)=\textrm{Tr}(B^\ell),\ \textrm{Tr}(t(A)^\ell)=\textrm{Tr}(t(B)^\ell)\ee
for $t$ given by (\ref{tform}). Adding $\textrm{Tr}(t(A+\mathbf{I})^\ell)=\textrm{Tr}(t(B+\mathbf{I})^\ell)$ condition, it was always sufficient for tested SRGs up to degree 29.

However, it is definitely not sufficient, what can be seen through its combinatorial interpretation: it calculates the number of such subgraphs in the graph. We could expand the graph by adding some number of new vertices in each edge, making it no longer able to recognize such structures. To handle it still in polynomial time we can for example include powers: test agreement for all $\alpha_1,\alpha_2,\alpha_3,\alpha_4,\alpha_5=1,\ldots,n$ of

$$\sum_{ij} (A^{\alpha_1})_{ai} (A^{\alpha_2})_{aj} (A^{\alpha_3})_{ij} (A^{\alpha_4})_{ib} (A^{\alpha_5})_{jb} $$
However, it is far from a formal proof, simplification is presented next hopefully to get closer.\\

\begin{figure}[t!]
    \centering
        \includegraphics{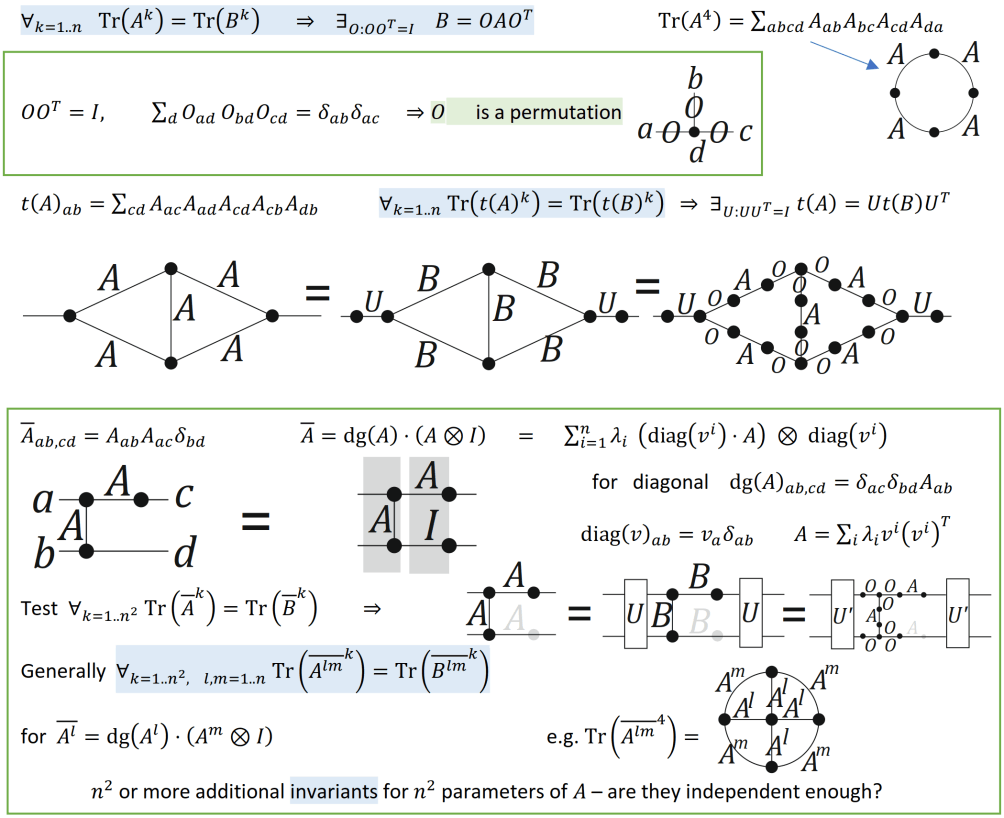}
        \caption{Analogous diagrams like in Fig. \ref{diag}, but this time to be directly applied to adjacency matrix to be used in edges of such graphs. Agreement of $\textrm{Tr}(A^k)$ for $k=1,\ldots,n$, corresponding to circular graphs with $A$ in edges and summation over verities, implies existence of orthogonal similarity matrix. We would like to additionally test agreement of invariants built from more general graphs, to restrict the space of possible similarity matrices to permutations only - getting sufficient condition for graph isomorphism. Using vertices of degree higher than 2, only permutation invariance is ensured - general orthogonal matrix can change them.
        For example for orthogonal matrix $O$, sum  $\sum_d O_{ad}O_{bd}O_{cd}=\delta_{ab}\delta_{ac}$ iff $O$ is a permutation. The task is to find invariants uniquely determining $n^2$ coefficients of $A$ up to a permutation: $O(n^2)$ invariants would be sufficient if being independent. We can easily build inexpensive families of such invariants like in Fig. \ref{diag}, the problem is getting a formal proof of sufficiency. Presented two families were tested to distinguish at least some strongly regular graphs. The first (central part of diagram) uses $t(A)$ deformation. Lower part of diagram presents approach with reduced the basic number of used matrices from 4 to 2 by working on tensor product. Beside algebraic, such invariants have also combinatorial interpretation: count the number of embeddings of its graph into the tested graph.}
       \label{graphs}
\end{figure}

\subsubsection{Simplifying by using tensor product}
There are $n^2$ coefficients in adjacency matrix (or less if using symmetry), hence $O(n^2)$ invariants uniquely determine them if only being independent enough. Like in Fig. \ref{graphs}, similarity invariants $\textrm{Tr}(A^k)$ can be seen as being defined by cycle-like graphs: with degree 2 vertices corresponding to summation of a given index appearing twice. Analogously using vertices of higher degree: with index appearing 3 or more times, we get permutation invariants, which are not necessary invariant for $A\to OAO^T$ transformation with $O$ not being a permutation. Such let say degree 3 vertex will get three $O$ matrices this way, and
\be OO^T=\textbf{I},\ \sum_d O_{ad}O_{bd}O_{cd}=\delta_{ab}\delta_{ac} \ \Rightarrow \ O \textrm{ is permutation}\label{perm}\ee
as $\sum_d (O_{ad})^2 =1$, hence sum of a higher power will be lower than 1, unless this is a canonical vector. Hence $O(n^2)$ or more such invariants with higher degrees would uniquely determine $A$ up to permutation if only being independent enough - the question is how to formally prove such independence.

While we could consider families of such graphs defining invariants, like in Fig. \ref{diag}, to get closer to such proof we should focus on the simplest sufficient family. The $t(A)$ basic construction before has used matrix $A$ four times, let us reduce it to two times using $\overline{A}$ construction visualized in Fig. \ref{graphs}. It was tested that it still distinguishes at least some strongly regular graphs, constructions using single appearance were not able to do it.\\

Denote $\overline{A}=\overline{A^{11}}$ for $n^2\times n^2$ matrix:
\be \overline{A^{lm}}_{ab,cd}=(A^l)_{ab} (A^m)_{ac} \delta_{bd}\ee
It can be represented as a tensor product:
\be \overline{A^{lm}}=\textrm{dg}(A^l)\cdot (A^m \otimes \textbf{I}_n)\quad\textrm{for}\quad \textrm{dg}(A)_{ab,cd}=A_{ab}\delta_{ac}\delta_{bd}\ee

Using eigendecomposition $A=\sum_{i=1}^n \lambda_i v^i (v^i)^T$ (not unique), we can get its tensor decomposition:
\be  \overline{A^{lm}}=\sum_{i=1}^n (\lambda_i)^l\ (\textrm{diag}(v^i)\cdot A^m) \otimes \textrm{diag}(v^i)= \ee
$$=\sum_{i,j=1}^n (\lambda_i)^l\, (\lambda_j)^m\ \left(\textrm{diag}(v^i)\cdot v^j(v^j)^T\right) \otimes \textrm{diag}(v^i)$$
for $\textrm{diag}(v)_{ab}=v_a \delta_{ab}$ diagonal matrix with (eigen)vectors on the diagonal.\\

In polynomial time we can test for example $$\forall_{k=1..n}\ \textrm{Tr}\left(A^k\right)=\textrm{Tr}\left(B^k\right),$$
\be \forall_{l,m=1..n}\,\forall_{k=1..n^2}\ \textrm{Tr}\left(\overline{A^{lm}}^k\right)=\textrm{Tr}\left(\overline{B^{lm}}^k\right)\ee
ensuring existence of orthogonal $n\times n$ matrix $OO^T=\textbf{I}_n$ and $n^2\times n^2$ matrices $U_{lm} (U_{lm})^T=\textbf{I}_{n^2}$ such that
$$B=OAO^T\qquad \forall_{l,m=1..n}\ \overline{A^{lm}} = U_{lm}\, \overline{B^{lm}}\, (U_{lm})^T$$
Substituting $B=OAO^T$ to the latter (tensor) equations, for sufficiency of graph isomorphism test we need to conclude that $O$ is a permutation.\\

While the proof is still missing, this form is relatively simple to work on, here are some remarks:
\begin{itemize}
  \item Substituting $B=OAO^T$, as in Fig. \ref{graphs} there appears higher order degree vertices surrounded by $O$ matrix, suggesting to use (\ref{perm}) observation which allows to conclude permutation.
  \item Instead of matrix $U$, there can be used some other consequence of confirmed similarity, like direct equality of traces of powers, agreement of characteristic polynomials, or through understanding consequences of eigenequation for $\overline{A}$.
  \item For strongly regular graphs we can focus on projection to one of 2 degenerated subspaces, what makes it sufficient to use only the $l=m=1$ case.
  \item It might be worth to add third $A$ in the bottom for symmetry - grayed in Fig. \ref{graphs}, tested that it also distinguishes at least some strongly regular graphs.
  \item We can also try to use combinatorial interpretation of $\textrm{Tr}\left(\overline{A^{lm}}^k\right)$: as number of such structures (subgraph isomorphisms) - vertices in distance $l$ from some central vertex, forming cycle with $m$ steps between each vertex. Above test ensures that for any length ($k$) of such cycle, their number is equal for both graphs.
  \item Performing summation over only the first index (instead of two in $\textrm{Tr}\left(\overline{A^{lm}}^k\right)$), we can choose this central point - getting vertex invariants, allowing e.g. to restrict the space of automorphisms.
  \item Alternative view on $\textrm{Tr}\left(\overline{A^{lm}}^k\right)$ is starting with $\textrm{Tr}\left(A^{m k}\right)$ cycle and adding $k$ "spokes" attached to the central point and every $m$ nodes of the cycle.
\end{itemize}
The question remains if such agreement is sufficient to conclude permutation (graph isomorphism), how to formally prove it?
\subsection{Sketch of proof for hypothetical test}
\begin{figure}[b!]
    \centering
        \includegraphics{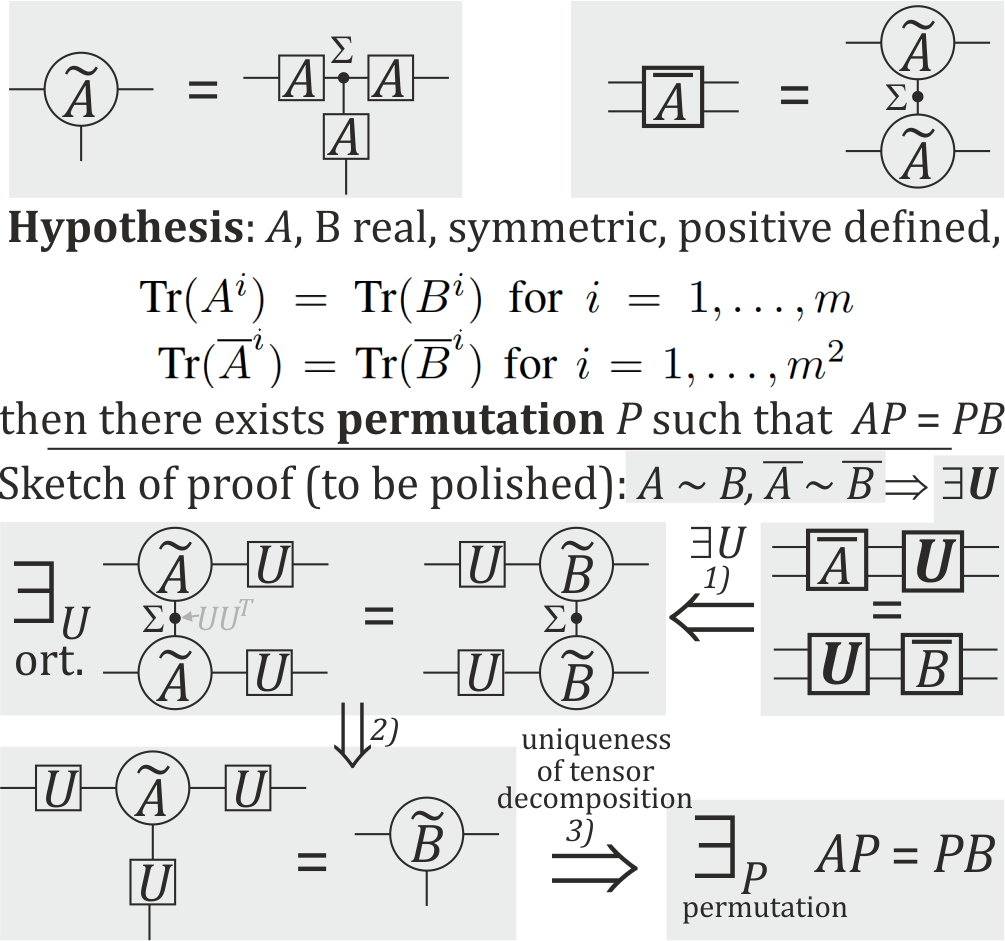}
        \caption{\textbf{Top}: construction of used tri-linear $\tilde{A}$, and $m^2 \times m^2$ matrix $\overline{A}$. \textbf{Middle}: hypothesis to be proven - that testing both similarities: $A\sim B$ and $\overline{A} \sim \overline{B}$, allows to imply existence of permutation as similarity matrix. If completing the proof, it would give polynomial time graph isomorphism test by using $A,B$ as adjacency matrices with added $mI$ for positive definiteness. \textbf{Bottom}: sketch of proof to be polished - details in text. }
       \label{sketch}
\end{figure}
This subsection contains attempt to prove being sufficient condition for graph isomorphism for type of tests as in the previous subsection, visualized in Fig. \ref{sketch} - to be polished into formal proof, maybe requiring some extension first.

Assume $A,B$ are symmetric $m\times m$ matrices, for convenience also positive defined. For graph isomorphism problem we can use $A,B$ as adjacency matrix plus  $mI$ to make them positive defined. If satisfying the proposed test of polynomial cost, we would like to prove existence of permutation $AP=BP$, which also defines isomorphism between the two graphs.

Part of this test is standard similarity test: $\textrm{Tr}(A^i)=\textrm{Tr}(B^i)$ for $i=1,\ldots,m$, ensuring existence of orthogonal matrix $O$ such that $AO=OB$. The symmetry assumption allows to decompose them using orthogonal matrices $O_A,O_B$ and the same diagonal matrix $D$ with positive eigenvalues:
\be A=O_A^T D O_A\qquad \qquad B=O_B^T D O_B \label{decomp} \ee

For additional test restricting orthogonal matrices to permutations, we will use the following symmetric tensor constructions: of tri-linear functions $\tilde{A}, \tilde{B}$ and $m^2 \times m^2$ matrix  $\overline{A}, \overline{B}$:
$$ \tilde{A}_{abc}:=\sum_{r=1}^m A_{ra} A_{rb} B_{rc} \qquad \tilde{B}_{abc}:=\sum_{r=1}^m B_{ra} B_{rb} B_{rc}$$
\be \overline{A}_{ab,cd}=\sum_i \tilde{A}_{aic} \tilde{A}_{bid}
\qquad \overline{B}_{ab,cd}=\sum_i \tilde{B}_{aic} \tilde{B}_{bid}\ee

We will sketch a proof (to be polished or add more tests) of the following:

\textbf{Hypothesis}: For $A,B$ real, symmetric, positive defined $m\times m$ matrices, if $\textrm{Tr}(A^i)=\textrm{Tr}(B^i)$ for $i=1,\ldots,m$ and $\textrm{Tr}(\overline{A}^i)=\textrm{Tr}(\overline{B}^i)$ for $i=1,\ldots,m^2$, then there exists similarity matrix  $P$ which is permutation: $PA=BP$.

The proof seems to require 3 steps visualized in Fig. \ref{sketch}:

\subsubsection{$m^2 \times m^2$ matrix $\textbf{U}$ decomposed into $m \times m$ matrices $U$}
Satisfaction of $\textrm{Tr}(\overline{A}^i)=\textrm{Tr}(\overline{B}^i)$ for $i=1,\ldots,m^2$ test ensures $\overline{A}\sim \overline{B}$ similarity: existence of
\be m^2 \times m^2\textrm{ orthogonal } \textbf{U}:\quad \overline{A} \textbf{U}= \textbf{U} \overline{B}\ee

As these are real symmetric matrices, we can diagonalize them, allowing to see this similarity as existence of common eigenspectrum $\{\Lambda_i\}$ with the same multiplicities, allowing e.g. to write for $x\in \mathbb{R}^m$ vectors:
\be (x^T \otimes x^T) \cdot \overline{A}\cdot (x \otimes x) =\sum_{i} \Lambda_i \left(\sum_{jk} x_j x_k\, V^i_{jk}\right)^2 \label{eqq2}\ee
where $V^i$ are $m\times m$ matrices representing eigenvectors of $\overline{A}$. They can be chosen as orthonormal basis, what can be written as $\textrm{Tr}\left(V^i\, (V^j)^T \right)=\delta_{ij}$.

For a chosen eigenvalue $\Lambda$,  we have eigenspace $E_\Lambda$ spanned by $D_\Lambda\subset \{1,\ldots,m\}$ subset of eigenvectors:
$$E_\Lambda = \left\{\sum_{i\in D_{\Lambda}} \alpha_i V^i: \alpha_i \in \mathbb{R}\right\}$$

For example transposing (\ref{eqq2}), also $E_{\Lambda}^T=\{V^T:V\in E_\Lambda\}$ is such eigenspace, hence they have to be equal $E_{\Lambda} = E_{\Lambda}^T$. Each such symmetric matrix can be decomposed $V=\sum_i \lambda_i\, v^i (v^i)^T$ for orthonormal vectors $\{v^i\}$, we know dimensionality of each subspace - suggesting (to be proven) that we can represent the eigenspaces as:
$$E_\Lambda = \left\{\sum_{i\in D_{\Lambda}} \alpha_i\, v^i\otimes v^i: \alpha_i \in \mathbb{R}\right\}$$
Building matrix from all $\{v^i\}$ vectors, and analogously for $\overline{B}$, we get a sketch of proof of existence of $U$ orthogonal $m\times m$ matrix such that:
\be \overline{A}\cdot (U\otimes U) = (U\otimes U) \cdot \overline{B} \label{eqq3}\ee
\subsubsection{Concluding $\tilde{A}\sim \tilde{B}$ similarity} from (\ref{eqq3}) type of similarity, the next step is to conclude $\tilde{A}\sim \tilde{B}$, remembering that $\overline{A}, \overline{B}$ are built of two copies of $\tilde{A}, \tilde{B}$, summed over common index. We can insert $U U^T=I$ in this summation, e.g.:
$$\overline{A}_{ij,kl}=\sum_{abc} \tilde{A}_{iak} U_{ab} U_{bc} \tilde{A}_{jcl}$$
In other words, from 4-index similarity (\ref{eqq3}), we would like to conclude 3-index similarity with two copies and summation:
\be \tilde{B}_{abc}=\sum_{ijk} \tilde{A}_{ijk} U_{ia} U_{jb} U_{kc}=\sum_{r=1}^m (AU)_{ra} (AU)_{rb} (AU)_{rc} \label{tril} \ee
This part of proof is missing in this version, a constructive one seems possible.
\subsubsection{Conclude permutation from uniqueness of tensor decomposition} For graph isomorphism problem, we need to propose additional tests, strengthening criterion to restrict from orthogonal similarity matrices to permutations. We would like to use tri-linear functions (tensor) for this purpose here:
\be\tilde{A}_{abc}:=\sum_{r=1}^m A_{ra} A_{rb} A_{rc} \qquad \tilde{B}_{abc}:=\sum_{r=1}^m B_{ra} B_{rb} B_{rc}\ee
Concluding analogy of their similarity (\ref{tril}), we could further use Theorem of uniqueness of tensor decomposition (\cite{tens,tens1}) if there is satisfied condition:
\be \textrm{uniqueness condition: }2m+2 \leq 3\, \textrm{rank}(A)\ee
The positive definiteness assumption implies $\textrm{rank}(A)=m$, hence this condition is satisfied. While we are using it for 3-linear function, we could analogously use it for 4-linear or higher (but as we could expect, 2-linear is insufficient for this condition).

Using this Theorem, we know there is unique tensor decomposition up to permutation of summed index in (\ref{tril}), allowing to conclude:
$$\sum_{r=1}^m B_{ra} B_{rb} B_{rc} = \sum_{r=1}^m (AU)_{ra} (AU)_{rb} (AU)_{rc} $$
\be \textrm{hence there exists permutation }P:\,PB=AU \ee
Multiplying $PB=AU$ by its transposition and using orthogonality of $U$, we get
$$PB^2 P^T=AU U^T A=A^2$$
As $A,B$ are positive defined, hence they are uniquely defined by $A^2, B^2$.  Hence we can conclude that $P$ is similarity matrix also for $PA=AB$ (hence the graphs are isomorphic).\\

\subsubsection{Isomorphism invariant polynomials}
The discussed sketch of proof in Fig. \ref{sketch} has weak point in 1): ensuring existence of $m^2 \times m^2$ orthogonal matrix, we would like to conclude existence of small $m \times m$ matrix. A natural direction to allow for such implication is trying to mark the two matrices, hoping to enforce that the $m^2 \times m^2$ rotation consists of two separate $m \times m$ rotations - internal for two matrices.

One way for distinguishing the two matrices is building polynomials with separate variables for them, extending on concept of characteristic polynomial, e.g.:
\be A_{\eta\mu} = \textrm{dg}(A)\cdot ((I+\eta A) \otimes (I+\mu A)) \ee
for $\textrm{dg}(A)_{ij,kl}=\delta_{ik} \delta_{jl} A_{ij}$ size $m^2 \times m^2$ diagonal matrix.

This way $A_{\eta\mu}$ is polynomial of $\eta,\mu$ variables. Building analogously $B_{\eta\mu}$ for adjacency matrix of the second graph, and testing if $\textrm{Tr}(A_{\eta\mu}^k)=\textrm{Tr}(B_{\eta\mu}^k)$ for all $k=1,\ldots,m^2$, what can be done in polynomial time, we get certainty of existence of large $m^2 \times m^2$ similarity matrices - this time depending on $\eta,\mu$ variables: existence of 2-parametric family of rotations.

\begin{figure}[t!]
    \centering
        \includegraphics{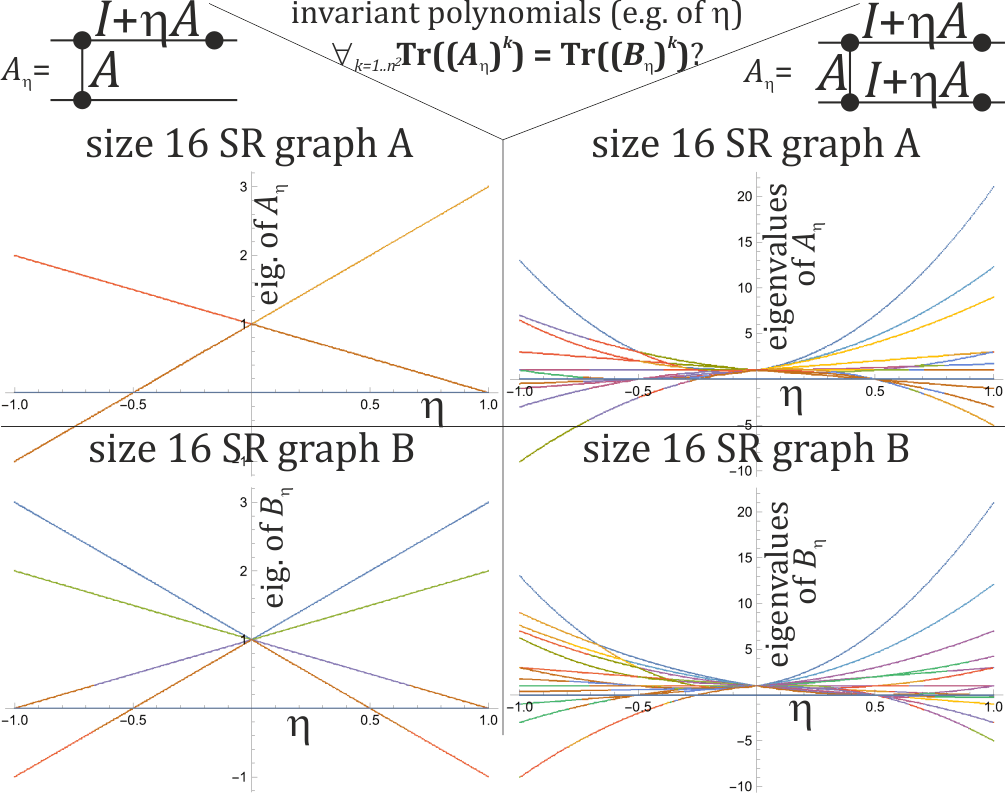}
        \caption{Two examples (left/right) of building isomorphism invariant polynomials (e.g. of $\eta$ here) for tests if traces of their powers for both graphs are equal (necessary for isomorphic). Bottom: comparison of eigenvalues of $A_\eta$, $B_\eta$ for two size 16 strongly regular graphs shown in Fig. \ref{srg} - their difference shows they are not isomorphic. We can see they are degree 1 or 2 polynomials of $\eta$. We can analogously extend to multiple variables, building graphs and using e.g. $I+\eta A + \mu \textbf{1}$ in edges. }
       \label{invpol}
\end{figure}

Unfortunately in this moment there is still no proof, but the direction seems promising, Fig. \ref{invpol} shows simpler 1-parameter $\eta=\mu$ polynomials distinguishing size 16 SRGs. We could analogously use larger graphs, more variables.

Beside $A$ matrix combinatorially corresponding to going to neighboring vertex, and $I$ identity matrix corresponding to staying in vertex, we can also use $\textbf{1}:=(1)_{ij}$ matrix made of constant values - corresponding to jumping to any vertex. Finally we can put e.g. $I+\eta A + \mu \textbf{1}$ type edges building invariant polynomials.

\subsection{Finding representation modulo permutation}
Let us consider here another approach. Having $m\times m$ adjacency matrix, set $\mathbb{A}:=\{P^TAP:P\textrm{ is permutation}\}$ has at most $m!$ matrices.

Two graphs are isomorphic iff such their sets of all permutations are identical. Finding a way to practically describe and compare such huge sets (at most size $m!$), we would get a graph isomorphism test.

Even more, defining a linear order on such huge set, two graphs are isomorphic iff they have identical minimal or maximal element of such order.

In other words, we could represent graph modulo isomorphism by taking vertex enumeration leading to adjacency matrix which is the smallest e.g. in some lexicographic order. Being able to efficiently calculate such representation, graph isomorphism could be tested by just checking if the representants of both graphs are equal.

To calculate them, it seems useful to use Sidon sequence $\{a_i\}$: such that all $a_i+a_j$ for $i>j$ pairs are different. Paul Erd{\"o}s and P{\'a}l Tur{\'a}n have shown~\cite{sidon} that for numbers up to $x$, there exists $(1-o(1))\sqrt{x}$ size Sidon set.

So assume $S=\{s_1,\ldots,s_m\}$ is some Sidon set realized using values bounded from above by $O(m^2)$, what is possible thanks to the above theorem. Define vector $$v :=(2^{s_1},\ldots,2^{s_m})^T$$
Observe that thanks to Sidon property, for 0/1 adjacency matrix $A$, value $v^T A v$ uniquely encodes this matrix - in binary representation given power can appear only zero or one time. It gives a chance for mathematically efficient search for minimum (or maximum) of $v^T P^T A Pv$ over permutation $P$.

Additionally, we can use Cholesky decomposition of $A + mI=LL^T$, where diagonal was added to make it positive defined. This way we get minimization/maximization of Euclidean norm of $\|L^T P v\|$ over permutations $\pi$:
\be \textrm{representant}(A)=\min_{\textrm{permutation }P}\ \|L^T P v\|\ee
Graphs are isomorphic iff have the same such representant.

\section{Conclusions}
The paper presented a few nonstandard reformulations on NP-complete problems: 3-SAT as the question of reaching zero of degree 4 nonnegative polynomial, plane or sphere crossing hypercube vertices problem, Subset Sum as integration problem, and Hamitlon cycles as power of matrix using Grassmann numbers or zeroing derivative. The P$\ne$NP assumption allows to localized the source of difficulty in these approaches, for example:
\begin{itemize}
  \item Algebraic view: the cost of calculating multivariate analogue of discriminant of degree 4 polynomial has to grow exponentially with the number of variables,
  \item Global optimization view: the number of local minima of the obtained polynomial has to grow exponentially with the problem size,
  \item Geometric view: the cost of determining if a plane or sphere intersects with $\{0,1\}^N$ grows exponentially with $N$,
  \item Integration problem: testing if $\int_0^{2\pi} \prod_i \cos(\varphi k_i) d\varphi$ is zero has exponential cost,
  \item Matrix representation of Grassmann numbers has to grow exponentially with their number.
\end{itemize}
Being able to falsify one of these statements would allow to conclude that P=NP.

While the P vs NP problem is usually attacked from the point of view of discrete mathematics, presented reformulations allow to take it to the field of abstract algebra, geometry, real analysis or continuous global optimization for better understanding of connections between these relatively far fields of mathematics, like complexity bounds, or finding some advanced approaches for solving or approximating NP-complete problems.\\

There were also presented two approaches for graph isomorphism problem which successfully distinguished tested strongly regular graphs, which seem the most difficult cases for discussed algebraic approaches. However, the proof that they will distinguish all non-isomorphic graphs still remains to bo found.

%
%
%

\bibliographystyle{IEEEtran}
\bibliography{cites}

\begin{thebibliography}{10}
\providecommand{\url}[1]{#1}
\csname url@samestyle\endcsname
\providecommand{\newblock}{\relax}
\providecommand{\bibinfo}[2]{#2}
\providecommand{\BIBentrySTDinterwordspacing}{\spaceskip=0pt\relax}
\providecommand{\BIBentryALTinterwordstretchfactor}{4}
\providecommand{\BIBentryALTinterwordspacing}{\spaceskip=\fontdimen2\font plus
\BIBentryALTinterwordstretchfactor\fontdimen3\font minus
  \fontdimen4\font\relax}
\providecommand{\BIBforeignlanguage}[2]{{%
\expandafter\ifx\csname l@#1\endcsname\relax
\typeout{** WARNING: IEEEtran.bst: No hyphenation pattern has been}%
\typeout{** loaded for the language `#1'. Using the pattern for}%
\typeout{** the default language instead.}%
\else
\language=\csname l@#1\endcsname
\fi
#2}}
\providecommand{\BIBdecl}{\relax}
\BIBdecl

\bibitem{cont}
Y.~Shang, M.~P. Fromherz, T.~Hogg, and W.~B. Jackson, ``Complexity of
  continuous, 3-sat-like constraint satisfaction problems,'' in \emph{IJCAI-01
  Workshop on Stochastic Search Algorithms}.\hskip 1em plus 0.5em minus
  0.4em\relax Citeseer, 2001.

\bibitem{min}
\BIBentryALTinterwordspacing
J.~Duda, ``3sat can be translated into continuous global optimization of
  polynomial,'' 2010. [Online]. Available:
  \url{https://groups.google.com/d/msg/comp.theory/OIEc2GK6JDg/eoz08rO2pLIJ}
\BIBentrySTDinterwordspacing

\bibitem{discr}
I.~M. Gelfand, M.~Kapranov, and A.~Zelevinsky, \emph{Discriminants, resultants,
  and multidimensional determinants}.\hskip 1em plus 0.5em minus 0.4em\relax
  Springer Science \& Business Media, 2008.

\bibitem{adiab}
D.~Aharonov, W.~Van~Dam, J.~Kempe, Z.~Landau, S.~Lloyd, and O.~Regev,
  ``Adiabatic quantum computation is equivalent to standard quantum
  computation,'' \emph{SIAM review}, vol.~50, no.~4, pp. 755--787, 2008.

\bibitem{LDPC}
R.~Gallager, ``Low-density parity-check codes,'' \emph{IRE Transactions on
  information theory}, vol.~8, no.~1, pp. 21--28, 1962.

\bibitem{grass}
F.~Berezin, ``The method of second quantization (pure and appl. physics, vol.
  24),'' \emph{New York}, 1966.

\bibitem{babai}
L.~Babai, ``Graph isomorphism in quasipolynomial time,'' in \emph{Proceedings
  of the 48th Annual ACM SIGACT Symposium on Theory of Computing}.\hskip 1em
  plus 0.5em minus 0.4em\relax ACM, 2016, pp. 684--697.

\bibitem{cameron}
P.~J. Cameron, ``Strongly regular graphs,'' \emph{Topics in Algebraic Graph
  Theory}, vol. 102, pp. 203--221, 2004.

\bibitem{shape}
\BIBentryALTinterwordspacing
J.~Duda, ``Normalized rotation shape descriptors and lossy compression of
  molecular shape,'' \emph{arXiv preprint arXiv:1509.09211}, 2015. [Online].
  Available: \url{https://arxiv.org/pdf/1509.09211}
\BIBentrySTDinterwordspacing

\bibitem{rotinv}
\BIBentryALTinterwordspacing
------, ``Polynomial-based rotation invariant features,'' \emph{arXiv preprint
  arXiv:1801.01058}, 2018. [Online]. Available:
  \url{https://arxiv.org/abs/1801.01058}
\BIBentrySTDinterwordspacing

\bibitem{tens}
J.~B. Kruskal, ``Three-way arrays: rank and uniqueness of trilinear
  decompositions, with application to arithmetic complexity and statistics,''
  \emph{Linear algebra and its applications}, vol.~18, no.~2, pp. 95--138,
  1977.

\bibitem{tens1}
A.~Stegeman, ``On uniqueness of the canonical tensor decomposition with some
  form of symmetry,'' \emph{SIAM journal on matrix analysis and applications},
  vol.~32, no.~2, pp. 561--583, 2011.

\bibitem{sidon}
P.~Erd{\"o}s and P.~Tur{\'a}n, ``On a problem of sidon in additive number
  theory, and on some related problems,'' \emph{Journal of the London
  Mathematical Society}, vol.~1, no.~4, pp. 212--215, 1941.

\end{thebibliography}
\end{document}